\documentclass[amsmath,amssymb,nofootinbib,notitlepage,superscriptaddress,twocolumn]{revtex4-2}
\pdfoutput=1

\usepackage[bottom]{footmisc}
\usepackage{enumitem}
\usepackage[colorlinks=true]{hyperref}
\usepackage[raggedright]{subfigure}
\usepackage{lipsum, babel}
\usepackage{comment}
\usepackage{slashbox}
\usepackage{amssymb}
\usepackage{tabularx}
\usepackage{cancel}
\usepackage{pifont}
\usepackage{soul}

\usepackage{mathtools,amsmath,amssymb,mathrsfs,color,esint}
\usepackage{array}
\usepackage{xcolor}
\newcolumntype{C}[1]{>{\centering\arraybackslash}p{#1}}
\newcommand{\p}{\partial}
\newcommand{\e}{\mathfrak{e}}

\begin{document}

\title{A Nonlinear Endpoint of Charged Horizon Instabilities}

\author{Zachary Gelles}
\email{zgelles@princeton.edu}
\affiliation{Department of Physics, Princeton University, Princeton, NJ 08540, USA}
\author{Frans Pretorius}
\email{fpretori@princeton.edu}
\affiliation{Department of Physics, Princeton University, Princeton, NJ 08540, USA}
    
\begin{abstract}
We numerically construct asymptotically extremal black holes through the nonlinear evolution of a charged scalar field. Our procedure --- which extends the work of Murata-Reall-Tanahashi \cite{murata_what_2013} to include charged scalar dynamics --- involves the fine-tuned scattering of wave packets within an initially super-extremal Reissner-Nordstr\"om spacetime. The resulting extremal solution develops an event horizon along which the energy density diverges and the charge density approaches a constant (i.e., the horizon forms with ``hair''). We investigate this behavior from the perspective of critical phenomena in gravitational collapse, giving evidence that dynamical extremal black holes act as universal threshold solutions modulo this family-dependent hair. As in the linear instability of fixed extremal backgrounds, the scalar field decays outside the dynamical extremal horizon. But just inside the horizon, the scalar curvature appears to develop unbounded growth. This implies that near-threshold solutions without a black hole could develop correspondingly large curvatures visible from future null infinity.
\end{abstract}

\maketitle

\section{Introduction}
Extremal black holes influence the dynamics of matter in their surroundings in a strikingly different manner than their sub-extremal counterparts. In particular, energy density decays more slowly outside an extremal black hole than it does outside a sub-extremal one. In fact, at the level of linear perturbations on a fixed background, the energy density of scalar field matter on an extremal horizon does not decay at all --- a result known as the Aretakis instability \cite{aretakis2011stability1,aretakis2011stability2,aretakis_kerr}.

Since Aretakis first established this instability, an important question has been the extent to which it remains when metric back-reaction is considered. One would expect --- if cosmic censorship is respected in these spacetimes --- that back-reaction will generically cause a sub-extremal black hole to form, limiting the time for which any Aretakis-like growth may occur. This was indeed found to be the case in the seminal paper of Murata-Reall-Tanahashi (MRT \cite{murata_what_2013}), who first addressed this problem by numerically solving the Einstein-Klein-Gordon system for scalar perturbations of an extremal Reissner-Nordstr\"om (RN) black hole in spherical symmetry. However, they also showed that for finely tuned data incident on an initially super-extremal spacetime, an extremal black hole can form at asymptotically late times. These ``dynamical extremal black holes" were found to exhibit gradient non-decay (and blow-up of higher-order derivatives) akin to that of the linearized Aretakis instability. This picture was then made mathematically rigorous in the work of Angelopoulos-Kehle-Unger (AKU \cite{angelopoulos2024nonlinear,AKU_2026}).

In this paper, we extend these results to the case of a \emph{charged} scalar field in spherical symmetry. Such an extension is interesting because electromagnetic interactions between the matter and the black hole are expected to enhance the instability. Indeed, Zimmerman \cite{zimmerman_horizon_2017} showed that on a fixed extremal RN background with a fixed electromagnetic potential, the energy density stored in a charged scalar field actually blows up along the horizon. Furthermore, analyzing the charged scalar field is valuable because it serves as a conceptual analog of the vacuum Kerr problem. We can think of charged scalar field perturbations of a charged black hole as a toy model for gravitational wave perturbations of a rotating black hole. And finally, the charged scalar field is important to understand in generality because it provides a matter model for which extremal black holes can form directly from collapse \cite{kehle_gravitational_2022}.

In our previous paper \cite{Gelles_charge}, we took the first step towards a fully nonlinear treatment of the charged scalar problem by evolving both the matter \emph{and} the electromagnetic potential on a fixed metric background. There, we found that the enhanced background instability of extremal RN under sufficiently charged scalar matter results in the accumulation of charge on the extremal horizon, imbuing the black hole with an extra strand of ``hair.'' 

In this paper, we add the next missing piece to the fully non-linear puzzle --- we solve the wave equation, Maxwell's equations, and Einstein's equations self-consistently in spherical symmetry. To explore the effect of Einstein's equations near extremality, we follow MRT \cite{murata_what_2013} and successfully construct dynamical extremal black holes by fine tuning charged wave packets incident on super-extremal spacetimes. Like MRT, we similarly find that enhanced Aretakis-like growth persists on the dynamical extremal horizon. We also find that more ``generic'' charged perturbations of initially extremal spacetimes lead to sub-extremal black holes, though we leave it to a follow-up work to describe the details of such scenarios. 

The fine-tuned nature of dynamical extremal spacetimes, which sit at a threshold of black hole formation, is emblematic of critical phenomena in the sense discovered
by Choptuik~\cite{choptuik1993universality}. If this correspondence is valid, then one expects the dynamical extremal solution to be universal in some sense. Universality would imply that near-threshold solutions follow identical scaling laws in parameter (moduli) space when fine tuned, independent of the particular one-parameter family of solutions used for the tuning. MRT conjectured, for example, that the surface gravity of near-threshold black holes scales as $\kappa\propto(M_i-M_*)^{1/2}$, where the initial super-extremal RN mass $M_i$ is the parameter tuned to the threshold value $M_*$

One important goal of our work is to begin investigating the correspondence between gravitational critical phenomena and dynamical extremal black hole formation in more detail. To this end, we show that the same near-threshold scaling solutions appear to emerge after tuning three distinct one-parameter families of initial data. Moreover, we find evidence consistent with the notion that the critical solution is ``one-mode unstable." We show that prior to the time when this putative unstable mode grows to relevance, the qualitative behavior of the solution does not depend on whether a black hole is actually present. However, we demonstrate at least a mild breaking of universality for this class of dynamical extremal black holes, as the particular value of the horizon charge density is family-dependent (as are the constants of proportionality governing Aretakis-like growth of gradients). 

We also show that the growth of the scalar field energy density becomes more pronounced in the interior region of dynamical extremal solutions, and we argue that this growth arises from what we call a ``blueshift focusing instability." Unlike the boundedness of the extremal horizon, we see trends suggesting unbounded growth of the Ricci scalar in the extremal interior. If this is a part of the putative universal critical solution --- and not an artifact of the initial RN singularity --- then arbitrarily large curvatures could be visible from future null infinity when tuned from the dispersive side of the threshold. 

The rest of this paper is organized as follows. In \S\ref{sec:numerics}, we present our numerical scheme, reviewing the formalism from our first paper \cite{Gelles_charge} and discussing how we incorporate the solution of Einstein's equations. We relegate some details of the initial data to Appendix~\ref{app:initdata}. In \S\ref{sec:results}, we present results of the fine tuning procedure, first restricting to a single one-parameter family of initial data and then expanding to three families. Throughout, we frame the discussion within the context of critical phenomena. In \S\ref{sec:sec_emerge}, we extend the analysis to the dispersive side of the threshold, and we argue that the emergence of divergent curvature gradients arises from a blueshift focusing property of (near-)critical threshold solutions. Some details of the latter calculation are left to Appendix~\ref{app:extremalredshift}. In \S\ref{sec:conclusion}, we conclude with a summary, and we discuss several important future steps to take to address remaining open questions.

\section{Numerical Scheme}
\label{sec:numerics}
Our numerical scheme is based on our previous work \cite{Gelles_charge}, except now we self-consistently incorporate metric back-reaction via Einstein's equations. We briefly summarize the matter model and numerical framework below. 

We work with a complex scalar field $\phi$ coupled to electromagnetism through a coupling constant $\e$. The real and imaginary parts of the scalar field are parameterized as\begin{align}
    \overline{\xi}\equiv {\rm Re}(r\phi),\quad \overline{\Pi}\equiv {\rm Im}(r\phi),
\end{align}
with $P\equiv |r\phi|$ the gauge-invariant modulus. The vector potential $A$ captures the evolution of the gauge-invariant enclosed charge $Q$, which is related to the radial electric field via Coulomb's law.

We employ double null coordinates $U$ (outgoing) and $V$ (ingoing) in spherical symmetry, for which the metric takes the form\begin{align}
    ds^2&=-2f\,dU\,dV+r^2d\Omega^2.
\end{align}
In these coordinates, the equations of motion for $\overline{\xi}, \overline{\Pi}, A_U,$  $A_V$, and $Q$ (all of which are functions of $U$ and $V$) are given in our previous paper \cite{Gelles_charge}. There, the metric functions $f(U,V)$ and $r(U,V)$ (the areal radius) are fixed to that of a background RN solution. Here, we view $f$ and $r$ as fully dynamical fields evolved via the Einstein equations. The equations of motion for these two fields are (cf. MRT \cite{murata_what_2013}):\begin{align}
    r_{,UV}&=-\frac{r_{,U}r_{,V}}{r}-\frac{f}{2r}\left(1-\frac{Q^2}{r^2}\right)
    \\
    f_{,UV}&=\frac{f_{,U}f_{,V}}{f}+\frac{f^2}{r^2}+\frac{2fr_{,U}r_{,V}}{r^2}-\frac{2f^2Q^2}{r^4}\\&-16\pi f\,{\rm Re}[(D_U\phi)(D_V\phi)^\star],
\end{align} 
which respectively come from the $UV$ and $\theta\theta$ components of Einstein's equations. Here, $D_\mu\phi$ is the gauged covariant derivative:\begin{align}
    D_\mu\phi\equiv \nabla_\mu\phi-i\e A_\mu\phi.
\end{align}

Initial data construction begins by identifying a point in the spacetime $(U_0,V_0)$, which corresponds to a two-sphere. To the causal past of this two-sphere, the spacetime is taken to be exact Reissner-Nordstr\"om with charge $Q_0$ and mass $M_0$. All dynamical fields are then evolved to the causal future of this two-sphere. This requires pasting initial data along a null cone, which we decompose into an ingoing hypersurface $\mathcal{N}_A$ and an outgoing hypersurface $\mathcal{N}_B$.

Along these hypersurfaces, $\overline{\xi}$ and $\overline{\Pi}$ are taken to be arbitrary functions, while $A_\mu$ and $Q$ are computed via Maxwell's equations, as in \cite{Gelles_charge}. Meanwhile, $r(U,V)$ is taken to match its Reissner-Nordstr\"om value in MRT gauge \cite{murata_what_2013}, and $f$ is computed by respectively solving the $UU$ and $VV$ components of Einstein's equations:\begin{align}
\label{eq:cons1}
   \mathcal{N}_A&:\,\, \left(\frac{r_{,U}}{f}\right)_{,U}+8\pi f^{-1}r(D_U\phi)(D_U\phi)^\star=0
   \\
   \label{eq:cons2}
  \mathcal{N}_B&:\,\,\left(\frac{r_{,V}}{f}\right)_{,V}+8\pi f^{-1}r(D_V\phi)(D_V\phi)^\star=0.
\end{align}
Solving these constraint equations in the presence of a dynamical metric requires care --- especially when the initial data is super-extremal --- and we describe the explicit solution procedure in Appendix~\ref{app:initdata}. 

\begin{figure}[h]
    \centering
    \includegraphics[width=0.4\textwidth]{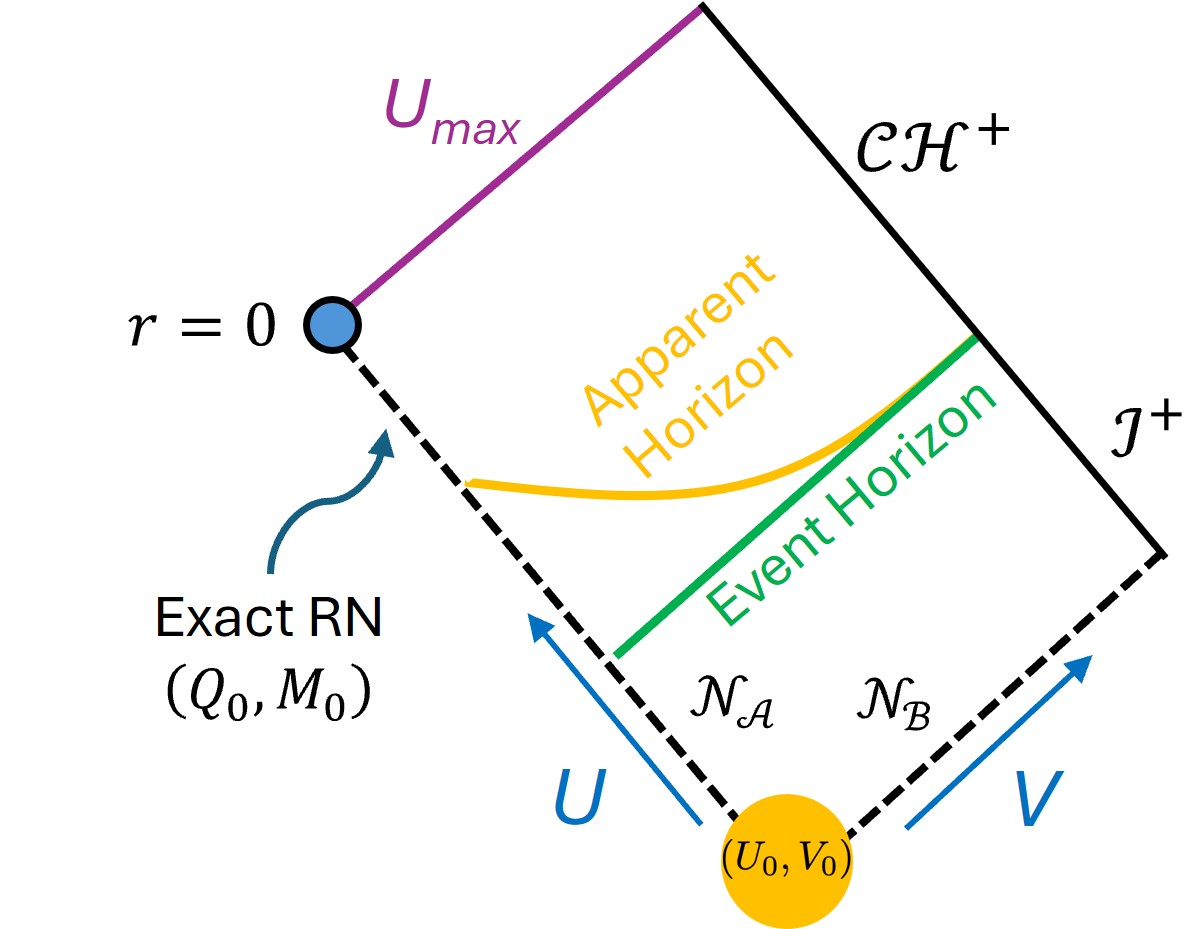}  
    \caption{Sample Penrose diagram of dynamical spacetime emanating from initial data pasted along the future null cone of the point $(U_0,V_0)$. When the matter perturbation is purely ingoing, the metric on $\mathcal{N}_A$ can be taken to match the exact RN solution, as depicted here. We evolve from $U_0$ to $U_{\rm max}$ along hypersurfaces of constant $U$, crossing any potential event/apparent horizons along the way.} 
    \label{fig:initdatafig}
\end{figure}

In this work, we will consider ingoing matter perturbations, which are supported only on $\mathcal{N}_B$. This allows us to set all metric variables to their exact RN values on $\mathcal{N}_A$.  Graphically, this initial data construction is depicted in Figure~\ref{fig:initdatafig}.

Once the initial data is constructed, we then use a finite-difference scheme to evolve along hypersurfaces of constant $U$ from $U=U_0$ to $U=U_{\rm max}$. The coordinate $U_{\rm max}$ is a Cauchy Horizon for evolution of our class of initial data, as it marks the time at which $\mathcal{N}_A$ intersects the RN singularity at $r=0$. 

As the data is evolved, we keep track of any apparent horizons, which are defined as surfaces across which the outgoing expansion $(\theta_+\propto r_{,V})$ flips sign. Extremal Reissner-Nordstr\"om contains just a single marginally trapped surface, across which $\theta_+$ reaches a local minimum at zero. In contrast, super-extremal RN has no apparent horizons at all. 

If an apparent horizon forms, the $U$ coordinate at which it ``intersects'' future null infinity on a Penrose diagram defines the event horizon of the spacetime, which separates outgoing null geodesics reaching null infinity from those who do not. These hypersurfaces are also depicted in Figure~\ref{fig:initdatafig}.

In the linear problem \cite{Gelles_charge}, we used static mesh refinement to cluster resolution around the event horizon. But the same procedure cannot be used in the non-linear setting, as the $U$-coordinate of the event horizon is unknown prior to evolution. Therefore, we instead implement a form of adaptive mesh refinement. In particular, we find that the numerical evolution converges most efficiently when we set our step size to be inversely proportional to the UV metric component at the outer boundary of the simulation: \begin{align}
\label{eq:refinement}
    \Delta U_{i+1}= \frac{\mathcal{C}}{f|_{V_{\rm max},i}}
\end{align} 
with $\mathcal{C}$ a constant. Such a criterion naturally clusters the grid cells around the event horizon, near which the double-null metric component grows large. Even when an event horizon is absent, Eq.~\ref{eq:refinement} still forces $\Delta U$ to shrink in the regions where the outgoing expansion becomes small and more resolution is needed. Eq.~\ref{eq:refinement} can be contrasted with different refinement criteria presented in Gundlach \& Martel \cite{Gundlach_Martel}, who also derive a formalism for numerical evolution in double-null coordinates. 

In analyzing the simulation results, we use the following expression to define a quasi-local (or renormalized Hawking) mass $M$ \cite{Poisson_massinflation,murata_what_2013} \begin{align}
    1-\frac{2M}{r}+\frac{Q^2}{r^2}=\nabla_\mu r\nabla^\mu r=-\frac{2r_{,U}r_{,V}}{f},
\end{align}
where all quantities in the above expression are functions of $(U,V)$. Unless otherwise specified, we will set the mass of the spacetime to unity at the origin of the initial data: $M(U_0,V_0)=M_0=1$, which will then grow in response to matter perturbations.

\section{Results}
\label{sec:results}
We first describe results from simulations where the scalar field initial data consists of a single, ingoing (i.e. supported only on $\mathcal{N}_B$), compact pulse of charged matter. In particular, we take the width of the pulse to be 20, and we take the shape of the pulse to match that used in our previous work \cite{Gelles_charge}:\begin{align}
     (\overline{\xi}+i\overline{\Pi})|_{\mathcal{N}_{A}}&=0
        \\
    (\overline{\xi}+i\overline{\Pi})|_{\mathcal{N}_{B}}&=\mathcal{A}(V)e^{-i\tilde{\omega} V}\label{eq:ampmod},  
\end{align}
where the envelope is given by\begin{align}
\label{eq:envelope}
    \mathcal{A}(V)&=\begin{cases}\mathcal{A}_0\exp\left[\frac{20}{4}
    \left(\frac{1}{V-20}-\frac{1}{V}\right)+1\right],&0\leq V\leq 20
    \\
    0,&{\rm else}.
    \end{cases}
\end{align}
Here, $\mathcal{A}_0$ sets the amplitude of the pulse, while $\tilde{\omega}$ modulates the envelope and can be interpreted as a local mass-to-charge ratio of the matter in the limit of large $r$ \cite{Gelles_charge}. For now, we will set $\mathcal{A}_0=0.01$ and $\tilde{\omega}=1$. The pulse begins at coordinates $(U_0,V_0)=(-1,0)$, where our choice of coordinate gauge then sets $r_{,U}=-1/2$ along $\mathcal{N}_A$. We will set the outer boundary of our simulation to $V_{\rm max}=400$, while the inner boundary is set based on our choice of coordinate gauge at $U_{\rm max}=2$. In practice, our code can typically evolve only to $U\approx 1.6$  (at which point $r=0.2$ on $\mathcal{N}_A$) before encountering numerical instabilities. Unless otherwise specified, the default simulation resolution will be $\Delta V=0.08$, while $\Delta U$ is set via Eq.~\ref{eq:refinement} with $\mathcal{C}=0.6$.

Additionally, as we saw in the linear problem \cite{Gelles_charge}, charged matter excites strongest instabilities on the extremal horizon when the coupling constant satisfies $\e Q_0\geq 1/2$. So throughout this work, we will fix the dimensionless coupling $\e Q_0=0.6$ unless otherwise specified.

Now, as we will detail in an associated paper, our simulations reveal that metric backreaction tends to \emph{decrease} the charge-to-mass ratio of the spacetime (see also \cite{murata_what_2013,hadar_near-extremal_2019,Gundlach_Martel}). In other words, perturbations of an extremal black hole will push the black hole towards sub-extremality, thus regulating any potential instabilities on the horizon. This is consistent with the numerical work of MRT \cite{murata_what_2013} and the analytic work of Hadar \cite{hadar_near-extremal_2019}, both of whom analyzed backreaction for an uncharged scalar field. 

However, it stands to reason that if the initial data is \emph{super}-extremal, then perhaps matter perturbations could decrease the charge-to-mass ratio of the spacetime until it becomes exactly unity. This is the crux of the ``dynamical extremal black holes" constructed in MRT using neutral scalar fields \cite{murata_what_2013}. Here, we will follow a similar procedure with charged scalar fields. 

\begin{figure*}[t]
    \centering
    \includegraphics[width=0.99\textwidth]{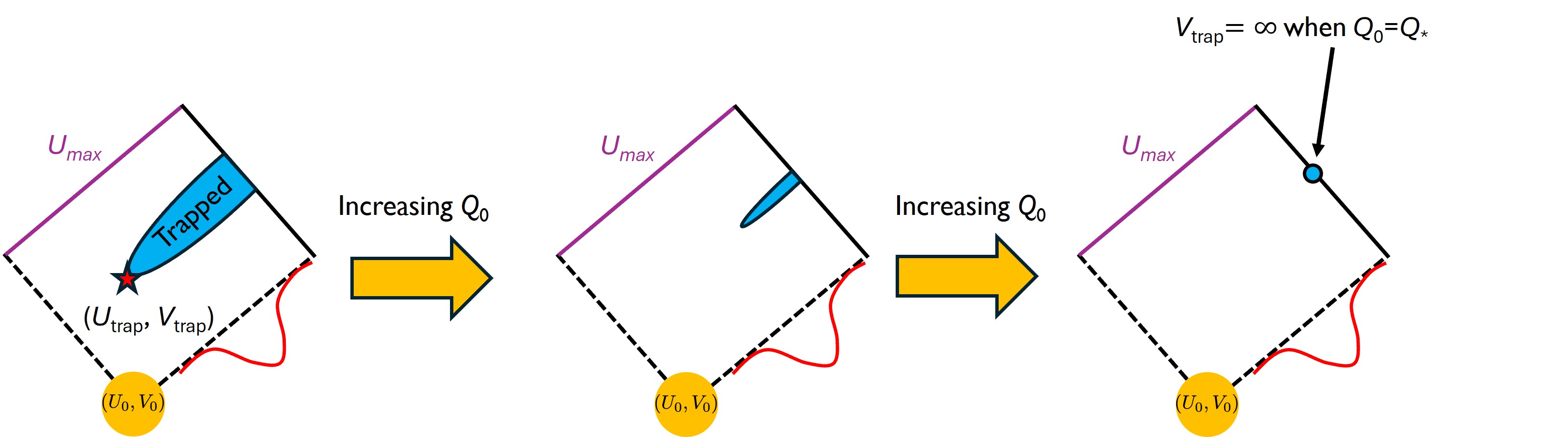}  
    \caption{Construction of a dynamical extremal black hole. Unlike Figure~\ref{fig:initdatafig}, here we show only the null diamond within which we evolve our initial data. What we call the background  spacetime ($V<V_0$) is super-extremal Reissner-Nordstr\"om with parameters $Q_0>M_0$. Given our ingoing pulse of charged scalar field (the red curve), a black hole forms for sufficiently small $Q_0$, and the trapped region exists for $V>V_{\rm trap}$. As the background charge $Q_0$ is increased to $Q_*$, the time $V_{\rm trap}$ at which a trapped region first appears is ostensibly pushed to $\infty$ (right panel); this fine-tuned case is the dynamical extremal black hole. For $Q_0>Q_*$ no trapped regions form in our evolution domain.}  
    \label{fig:dynERNform}
\end{figure*}

\subsection{The Extremal Threshold}
\label{sec:extremalthreshold}
To construct a dynamical extremal black hole, we set the metric on $\mathcal{N}_A$ to be super-extremal RN with parameters $Q_0>M_0$. Since the back-reaction of the scalar field will decrease the charge-to-mass ratio of the spacetime, then for a sufficiently large perturbation, a trapped region will form to the future of the initial data. We say that the trapped region ``appears" at coordinates $(U_{\rm trap},V_{\rm trap})$, meaning that $V_{\rm trap}$ is the smallest $V$ for which a trapped surface is ever present. The resultant spacetime will be a sub-extremal black hole.

But if we increase $Q_0$ while keeping $M_0$ and the pulse profile fixed, then the trapped region will shrink, and $V_{\rm trap}$ will grow. As we fine-tune $Q_0$, we can push $V_{\rm trap}\to\infty$, at which point the trapped region becomes arbitrarily small, and the inner/outer horizon bounding the trapped region degenerate to a single marginally trapped surface ``at'' future null infinity. In other words, an extremal black hole forms. We say that this dynamical extremal black hole forms at the critical threshold $Q_0=Q_*$. This process is shown pictorially in Figure~\ref{fig:dynERNform}, which matches onto the analogous construction of AKU \cite{angelopoulos2024nonlinear}.

With this conceptual picture in mind, we perform the fine tuning process numerically using a bisection search for $Q_*$. We will refer to spacetimes with $Q_0<Q_*$ as the ``BH-forming" side of the threshold, as they form black holes. And we will refer to spacetimes with $Q_0>Q_*$ as the ``dispersive" side of the threshold, where no trapped surfaces form within the domain of our numerical evolution. In this section, we focus on the BH-forming side of the threshold. The dispersive side will be discussed in \S\ref{sec:sec_emerge}. 

\begin{figure}[h]
    \centering
    \includegraphics[width=0.49\textwidth]{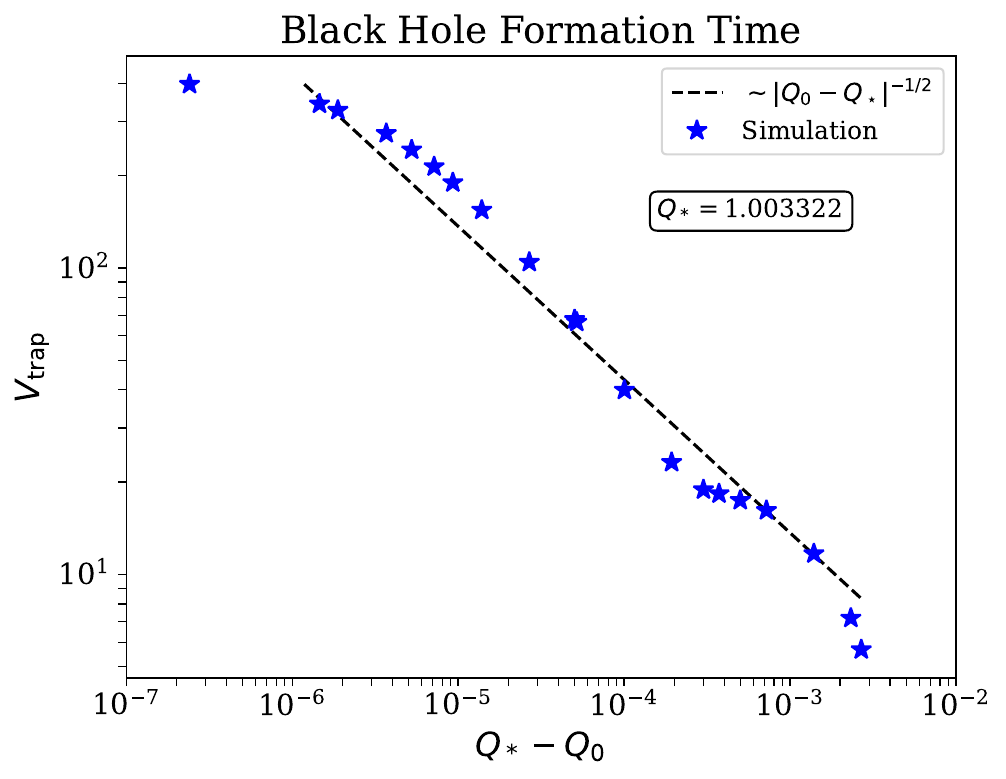}  
    \caption{Fine tuning of $Q_0$ to push $V_{\rm trap}$ --- the time at which a black hole forms --- to infinity. We find that $V_{\rm trap}\sim |Q_0-Q_*|^{-1/2}$ in this regime.} 
    \label{fig:Vtrapfig}
\end{figure}

For the particular family of initial data described above, the trends from evolution suggest $V_{\rm trap}\to\infty$ as $Q_0\to 1.0033218$, fine-tuned to 7 digits\footnote{We cannot fine-tune past 7 digits for this set of simulation runs, as further fine-tuning pushes $V_{\rm trap}$ out of the simulation domain. We additionally note that the value of $Q_*$ obtained through bisection depends on the simulation resolution, but we have confirmed that it is converging to a constant.}. We accordingly take $Q_* \approx 1.0033218$. This result is depicted in Figure~\ref{fig:Vtrapfig}.

As we approach the critical point, we attempt to identify the effective power-law $p$ with which $V_{\rm trap}$ diverges:
\begin{align}
\label{eq:Vtrapscaling}
    V_{\rm trap}\propto (Q_*-Q_0)^{-p}.
\end{align}
In general, fitting for $p$ is quite difficult, as it is highly ``degenerate" with $Q_*$ given the modest range in $\ln(Q_*-Q_0)$ we have been able to explore: a small variation in $Q_*$ can lead to a seemingly different power-law $p$. To break this apparent degeneracy, we will ultimately need to evolve simulations out to several more orders of magnitude in $V$, which is beyond the scope of our current numerical tools.

For now, we simply note that $p=1/2$ both reproduces the data cleanly and coheres well with our theoretical intuition. This fiducial $1/2$ scaling is shown as the black dashed line in Figure~\ref{fig:Vtrapfig}, matching onto the simulation results (blue stars) in an intermediate range $1\lesssim Q_0\lesssim Q_*$. When $Q_0$ is too close to $Q_*$, the reliableness of the power-law fit breaks down since the relative uncertainty in $Q_*-Q_0$ becomes large. Furthermore, when $Q_0$ is close to $1$ and $V_{\rm trap}$ enters the regime of the transient (recall that the pulse fired at the black hole has a width of $20$), the scatter around the power-law fit begins to grow as well.

A closely related $1/2$ scaling already appears for the neutral field in MRT \cite{murata_what_2013}, though they do not explicitly relate it to the quantity $V_{\rm trap}$. The same scaling was also seen in numerical simulations of extremal critical collapse in the Einstein-Maxwell-Vlasov system \cite{east2025gravitational}. And though perhaps unrelated, a similar $1/2$ conformal weight emerges from the near-horizon symmetry of near-extremal Kerr \cite{Gralla_Zimmerman}. Critical exponents of this nature have also been seen in a renormalization-group analysis of extremal solutions \cite{Yang_Renorm}.

The agreement between the simulation results and the $p=1/2$ scaling in Figure~\ref{fig:Vtrapfig} is thus well-founded. For completeness, however, we explore in Appendix~\ref{app:critfit} the possibility that $p$ deviates from $1/2$. There, we perform a more robust fit to the data, finding that $p\in[0.44,0.80]$ (with uncertainties capturing the potential transient effects in the simulation results). This is consistent with $1/2$ but affords the possibility that $V_{\rm trap}$ may scale more steeply in the near-threshold regime. Future work pushing to larger values of $V$ will be needed to constrain the exact value of $p$ to a higher degree of precision. 

\begin{figure}[h]
    \centering
    \includegraphics[width=0.5\textwidth]{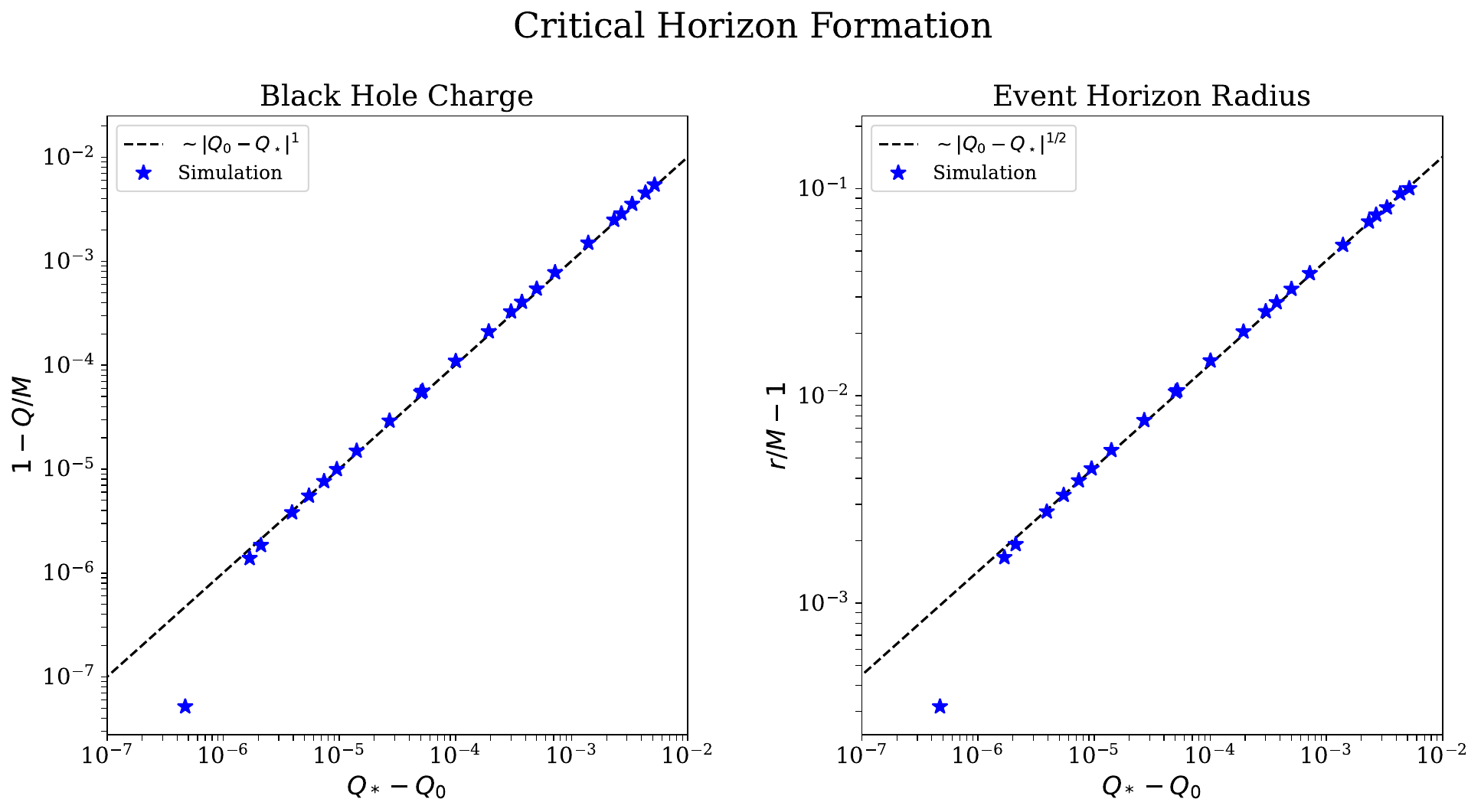}  
    \caption{Black hole properties as $Q_0\to Q_*$. Both the charge of the black hole and radius of its event horizon approach the black hole's mass, indicating that the threshold black hole is extremal. The charge of the black hole approaches extremality linearly, while the radius of the event horizon approaches extremality with the critical exponent of $1/2$. }
    \label{fig:criticalQ}
\end{figure}

Regardless, however, the trends from our simulations suggest that as $V_{\rm trap}\to \infty$, the horizon geometry approaches that of an extremal Reissner-Nordstr\"om black hole. We show this in Figure~\ref{fig:criticalQ}, where the radius $r$ and charge $Q$ of the resultant black hole approach extremality with the related power laws\begin{align}
\label{eq:BHscaling}
    1-Q/M&\propto Q_*-Q_0
    \\
    \label{eq:BHscaling2}
    r/M-1&\propto (Q_*-Q_0)^{1/2}.
\end{align}
Both quantities are numerically measured on an estimate of the event horizon that we simply take to be the apparent horizon on the largest $V$ in our simulation domain (in this case, $V_{\rm max}=400$). We note that in these simulations, the apparent horizons are always connected, in contrast to the discontinuous apparent horizons that appear in the formation of extremal black holes from direct collapse \cite{kehle_gravitational_2022}.  Furthermore, we have confirmed that the scaling exponents in Eqs.~\ref{eq:Vtrapscaling}-\ref{eq:BHscaling2} converge with simulation resolution. 

Like Figure~\ref{fig:Vtrapfig}, the agreement in Figure~\ref{fig:criticalQ} between our simulation results and the pure power-law breaks down as $Q_0\to Q_*$, where the relative uncertainty in $Q_*$ grows large. But the relations hold very well everywhere else, and unlike the scaling of $V_{\rm trap}$ (Eq.~\ref{eq:Vtrapscaling}) seem to be largely unaffected by transient dynamics far from threshold. This is at least in part due to the fact that Eqs.~\ref{eq:BHscaling}-\ref{eq:BHscaling2} are measured at late times (large $V$) in the evolution, and therefore include all accretion that has occurred from $V=V_{\rm trap}$ to $V=V_{\rm max}$; in other words, each point on the plots in Figure~\ref{fig:criticalQ} includes essentially the same integrated transient, leaving the far-from-threshold scaling unaffected for this class of initial data.


The fact that a scaling solution exists suggests that the threshold solution is universal in the sense of gravitational critical collapse. We explore this possibility in the next subsection.


\subsection{Universality}
\label{sec:universality}
If the dynamical extremal black hole is a universal solution, then both the threshold solution and the near-threshold scaling relations should hold regardless of the one-parameter family of initial data used to cross the respective critical points. In this section, we check for consistency with universality by looking at two additional, distinct one-parameter families.

\begin{figure}[h]
    \centering
    \includegraphics[width=0.49\textwidth]{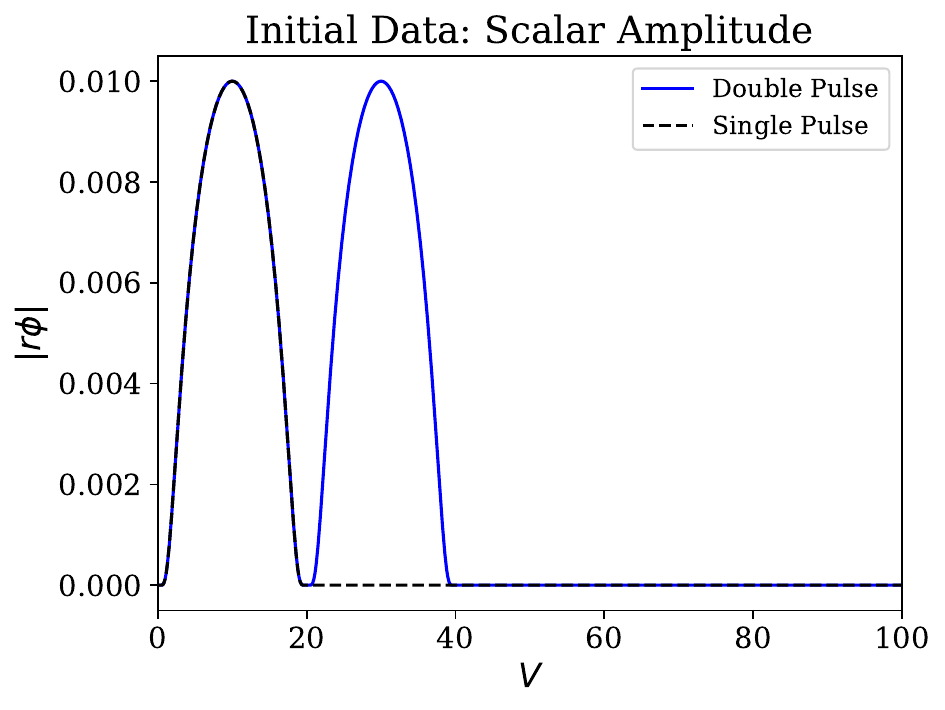}  
    \caption{Initial conditions for the scalar amplitude along $\mathcal{N}_B$. The ``single-pulse" --- employed in the previous subsection and shown as a black dashed line in this figure --- is presented in contrast to the ``double-pulse" in blue.} 
    \label{fig:doubleinitfig}
\end{figure}

First, we will show that the same critical exponent emerges after modifying our initial data profile to consist of a ``double-pulse": a compactly supported pulse from $V=0$ to $V=20$, followed by a subsequent compactly supported pulse from $V=20$ to $V=40$, as depicted in Figure~\ref{fig:doubleinitfig}. 

\begin{figure*}[t]
\centering
\begin{subfigure}
\centering
\includegraphics[width=0.99\textwidth]{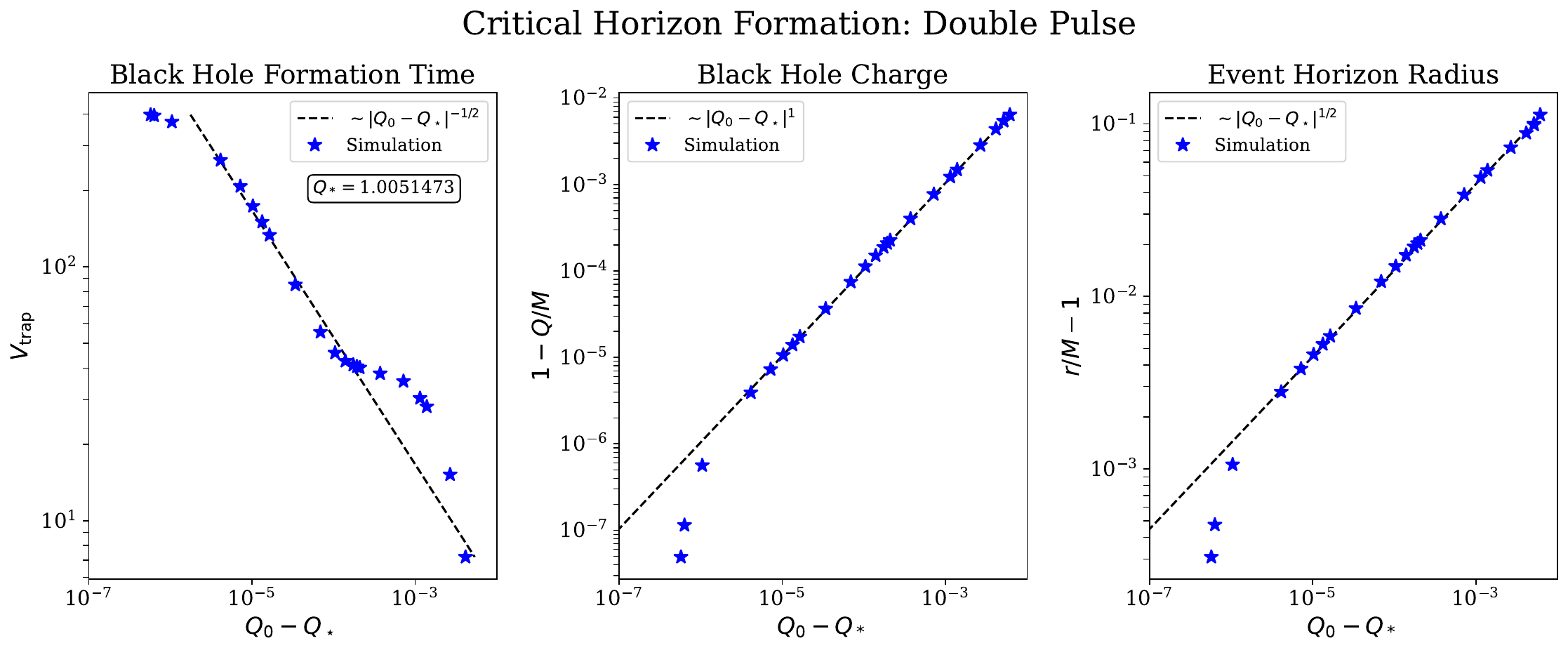}  
\end{subfigure}
\vspace{0.4cm}
\begin{subfigure}
\centering
 \includegraphics[width=0.99\textwidth]{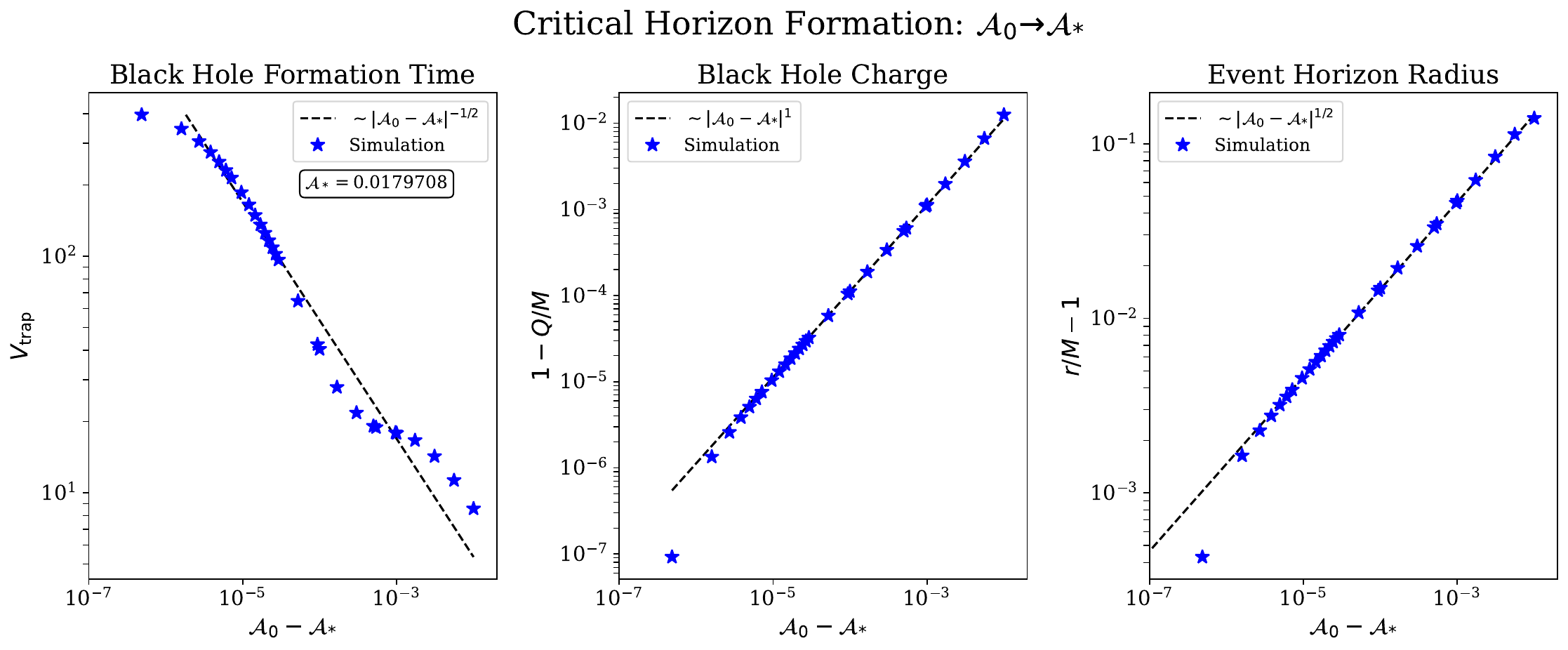}  
    \caption{Top: Black hole properties as $Q_0\to Q_*$ for the double-pulse. Bottom: Black hole properties as $\mathcal{A}_0\to \mathcal{A}_*$ for the single-pulse. In both cases, $V_{\rm trap}\to\infty$ as the threshold is approached, in agreement with Figure~\ref{fig:Vtrapfig}. This results in an extremal black hole with $r=Q=M$, as can be seen by extrapolating the trends shown in the central and right panels of each row.} 
    \label{fig:criticalamplitude}
\end{subfigure}

\label{fig:both}
\end{figure*}


Since this initial data is different from the single-pulse used in the previous subsection, it produces different transient dynamics. However, when we repeat the numerical fine-tuning procedure with the double-pulse data (here with $Q\rightarrow Q_{*,\rm double}\approx1.0051471$), we find the {\em same} scaling relations for $V_{\rm trap}$, $Q/M$, and $r/M$ as were found for the single pulse in Eqs.~\ref{eq:Vtrapscaling}-\ref{eq:BHscaling2}. These results are depicted in the top row of Figure~\ref{fig:criticalamplitude}, for which we see the power-law behavior emerging as we enter the critical regime. Note that since the double-pulse is twice as wide as the single pulse, $V_{\rm trap}$ deviates more substantially from the pure-power law at large $Q_*-Q_0$ than it did in Figure~\ref{fig:Vtrapfig}.


As a further check of consistency with universality, we revert back to the single pulse initial data, but now approach the critical point by varying $\mathcal{A}_0$ --- the amplitude of this pulse (see Eq.~\ref{eq:envelope}) --- while keeping $Q_0$ fixed. 
Arbitrarily choosing $Q_0=1.01$, we then find that the dynamical extremal black hole can be created in the limit $\mathcal{A}_0\rightarrow\mathcal{A}_*\approx0.0179708$ (note here that black holes form for $\mathcal{A}_0>\mathcal{A}_*$) . As we approach $\mathcal{A}_*$, the data are visually consistent with the same scaling properties: \begin{align}
    V_{\rm trap}&\propto (\mathcal{A}_0-\mathcal{A}_*)^{-1/2}
    \\
    r/M-1&\propto (\mathcal{A}_0-\mathcal{A}_*)^{1/2}
    \\
    1-Q/M&\propto (\mathcal{A}_0-\mathcal{A}_*)^{1}.
\end{align}
These scalings are depicted in the bottom row of Figure~\ref{fig:criticalamplitude}, and the uncertainties in the exponents are discussed in Appendix~\ref{app:critfit}. Note that in all these scaling relationships, the constants of proportionality can be obtained from fits to the data if desired. However, these constants are family-dependent and are certainly not universal.


While we have explored the threshold of only three sets of initial data families, the persistence of the same critical exponent suggests that the dynamical extremal black hole may be universal in a similar sense to threshold solutions in gravitational critical collapse \cite{choptuik1993universality,Hara_1996,Hod_1997_crit,garfinkle1998scaling,gundlach2007critical,Gundlach:2025yje}. In gravitational critical collapse, there exists a universal threshold solution that is ``one-mode unstable,'' and it is the eventual growth of this putative mode that leads to either black hole formation or ``dispersal'' (e.g. \cite{gundlach2007critical}). The fine-tuning process in our numerical evolution therefore effectively tunes the amplitude of this mode to zero. 

However, there are several conceptual difficulties that prevent this picture from applying verbatim to dynamical extremal black holes. The first difficulty comes from the existence of the exact extremal RN solution, which seems to have meaningfully different structure than the dynamical extremal black holes. The latter exhibit a marginally trapped surface only at $V=\infty$, whereas static extremal RN exhibits a marginally trapped surface spanning all $V$. Moreover, dynamical extremal black holes contain scalar field energy throughout the relevant region of the spacetime.
From AKU \cite{angelopoulos2024nonlinear}, we expect there to be a higher-dimensional family of dynamical extremal black holes --- perhaps all ``one-mode unstable" --- with the static extremal RN solution being the vacuum limiting case. 

A second difficulty is that generic perturbations of extremal RN lead to sub-extremal black holes \cite{murata_what_2013,hadar_near-extremal_2019}, whose interiors --- in contrast to their exteriors --- are unstable \cite{Penrose_blueshift,Poisson_massinflation,Dafermos:2003vim,Marolf_Ori,luk2019strong,van2018stability}. To what extent, if any, do dynamical extremal black holes exhibit similar instabilities in their interiors?  For a near-threshold solution that eventually forms a black hole, the notion of ``interior" naturally corresponds to the region inside the event horizon. For near-threshold solutions that do \emph{not} form a black hole, we instead introduce the concept of a ``would-be interior": the region that would otherwise have been trapped if a black hole had formed with the same amplitude but opposite-sign growing mode. If dynamical extremal black holes are universal in the above sense, then the near-threshold solutions evaluated prior to $V_{\rm trap}$ (and after transients have dispersed/decayed) should be independent of the sign of the growing mode. 

A third puzzle --- related to the above interior issues --- concerns what happens to all decaying modes of the critical solution. In scalar field collapse, for example, all the decaying modes are effectively radiated away, leaving behind the universal, self-similar collapsing core. But on the approach to dynamical extremal RN, it is unclear what happens to the modes propagating into the interior (or inner core) of the solution. If these interior modes are effectively ``trapped'' until dispersal (or truly trapped in the case of black hole formation), universality cannot hold arbitrarily far into the interior. This begs the question: In what vicinity of the dynamical extremal horizon is the spacetime described by a universal solution? 


We are not able to definitively resolve any of the above conceptual difficulties. But in the following section, we will try to shed some light on the nature of the instabilities that emerge as we approach the threshold solution.

\section{Emergent Instability}
\label{sec:sec_emerge}

In this section we will demonstrate that the extremal threshold solution exhibits behavior akin to the Aretakis instability of extremal RN. In particular, we will see that this behavior emerges in near-threshold solutions on \emph{either} side of the critical point, indicating that Aretakis-like growth can occur even in the absence of a black hole.

Extremal black hole interiors have been the subject of interest for well over a decade. The works of Marolf \cite{Marolf_extremal} and Garfinkle \cite{Garfinkle_extreme} conjectured that extremal black hole interiors should be unstable. MRT \cite{murata_what_2013} then confirmed numerically that parts of the Marolf-Ori sub-extremal shock \cite{Marolf_Ori} persist in the extremal limit, where observers crossing into the interior experience asymptotically divergent radial acceleration. 

In this section, we will further present evidence of gradient blow-up in both the interiors of near-threshold solutions containing a black hole, as well as the would-be interiors of near-threshold solutions lacking a black hole. We will explain such phenomena by qualitatively synthesizing the arguments of previous papers into a ``near-extremal focusing instability," wherein the focusing of outgoing radiation causes the gradient-blow up both on the horizon and in the interior.

Finally, we will speculate on whether this focusing instability could create a curvature singularity --- and whether such a singularity formed on the dispersive side of the critical point could be visible from future null infinity.

\subsection{Horizon Instability: Aretakis}
\label{sec:aretakis}
Extremal horizons (both RN and Kerr) are known to be unstable to perturbations of a scalar field when the background metric is held fixed. This instability --- originally formulated by Aretakis \cite{aretakis2011stability1,aretakis2011stability2,aretakis_kerr} --- emerges due to the lack of redshift on the extremal horizon. 

For electrically neutral scalar fields, the energy density $\rho_E\sim (\partial_r\phi)^2$ stored in the scalar field remains constant along the extremal horizon, with higher-order derivatives blowing up in $V$. This result was numerically shown to persist in the non-linear setting by MRT \cite{murata_what_2013}, and more recently analytically shown by \cite{porfyriadis2025approaching}: dynamical extremal black holes formed from neutral scalar fields exhibit late-time dynamics akin to the linear Aretakis instability.

The Aretakis instability has been shown to become enhanced in the linear setting when the scalar field is charged. Indeed, Zimmerman \cite{zimmerman_horizon_2017} found that when the background metric and electromagnetic potential are both fixed, the energy density stored in the scalar field $\rho_E\sim (\partial_r|\phi|)^2$ actually \emph{diverges}, with analogous blowup results holding for the Kerr problem \cite{casals_horizon_2016}. And in our previous work \cite{Gelles_charge}, we incorporated a dynamical Maxwell field (though still on a fixed background metric), finding that a constant, nonzero charge density $\rho_Q\sim \p_r Q$ emerges on the extremal horizon as well; in other words, charge accumulates along the extremal horizon without falling into the black hole or escaping off to null infinity.

But up until now, no one has addressed the fully nonlinear endpoint of the charged Aretakis instability. Can energy density really blow up along an event horizon if the metric backreaction of Einstein's equations is incorporated self-consistently?

Our numerical simulations indicate that the answer is yes. To show this, we extract the energy density and charge density from our simulations along the black hole's outer horizon as we fine tune $Q_0\to Q_*$. Unless otherwise specified, in this section we use initial data consisting of the single-pulse setup employed in \S\ref{sec:extremalthreshold}. In the dynamical spacetime, we take $\p_r Q$ as a proxy for the charge density (charge per unit volume), and $T_{rr}$ as a proxy for the energy density (energy per unit volume). In terms of the gauge-invariant matter fields $P\equiv |r\phi|$ and $Q$, this stress tensor component can be expressed as \cite{Gelles_charge}\begin{align}
    T_{rr}= 2\left[\left(\frac{Q_{,r}}{8\pi \e rP}\right)^2+\left(\left(\frac{P}{r}\right)_{,r}\right)^2\right].
\end{align}
Here and throughout the rest of this paper, commas denote partial differentiation, and radial derivatives are evaluated along contours of constant $V$. We numerically estimate the $U$ coordinate of the event horizon by identifying $U_{\rm EH}$ with the smallest value of $U$ along which a trapped surface appears in the simulation domain.

\begin{figure}[h]
    \centering
    \includegraphics[width=0.49\textwidth]{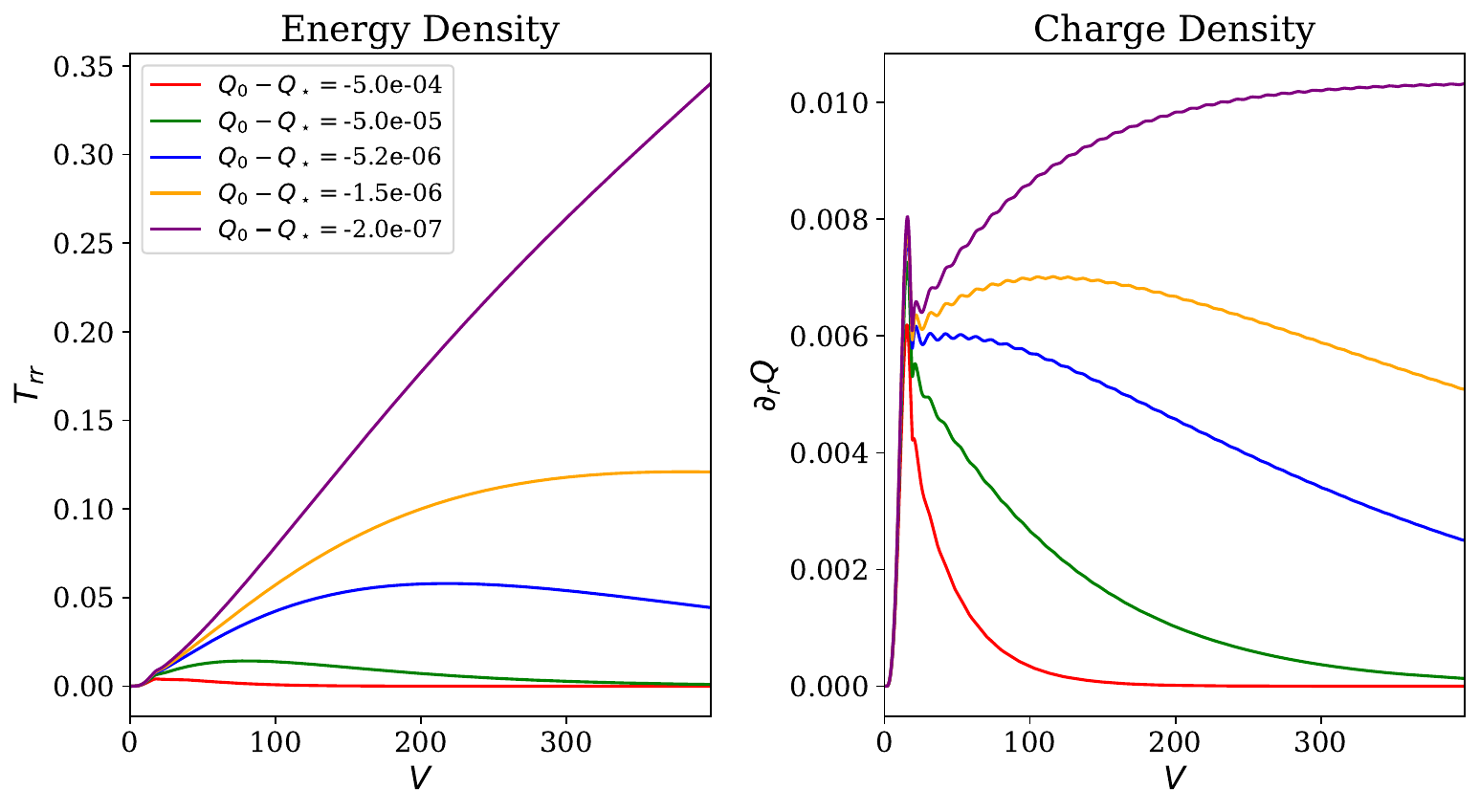}  
    \caption{Energy density and charge density along the event horizon of a black hole as $Q_0\to Q_*$ and the black hole approaches extremality. Extrapolating the trends, the energy density grows without bound on the dynamical extremal horizon, while the charge density approaches a constant.} 
    \label{fig:horizongrowth}
\end{figure}

In Figure~\ref{fig:horizongrowth}, we plot the energy density and charge density along this event horizon proxy as we tune $Q_0\to Q_*$ for the same set of initial data used in \S\ref{sec:extremalthreshold}. In this figure, we see that as extremality is approached, we observe growth of the energy density and asymptotic constancy of the charge density along the extremal horizon; this suggests that the results of our previous work \cite{Gelles_charge} may carry over to the non-linear problem. Later in this section, we will show that the time at which the energy density stops growing scales as $|Q_0-Q_*|^{-1/2}$ --- consistent with a genuine blow-up in the appropriate limit.

\begin{figure}[h]
    \centering
    \includegraphics[width=0.45\textwidth]{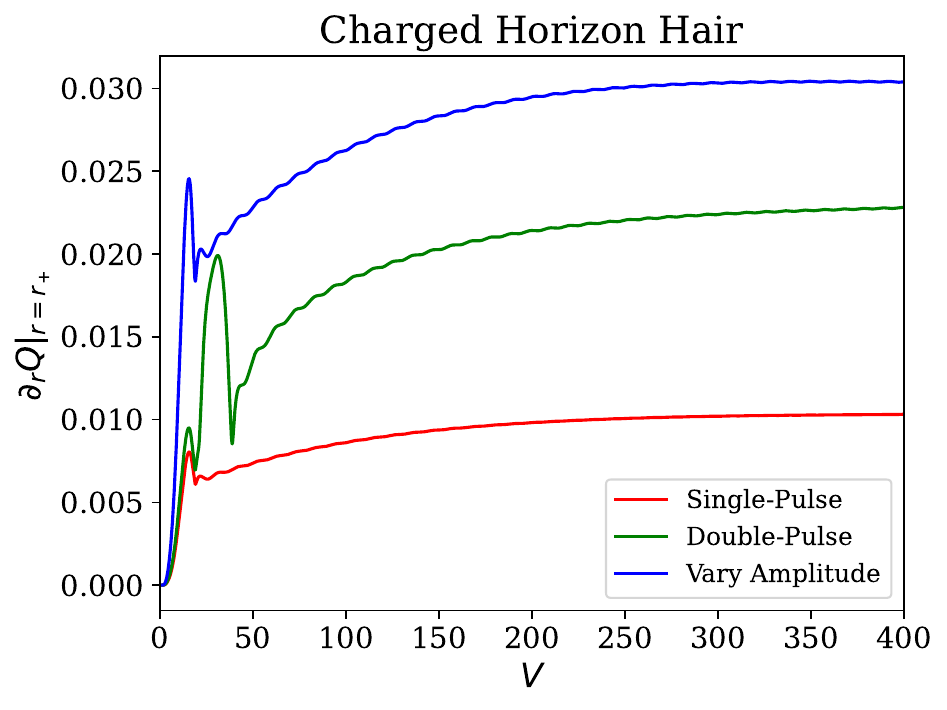}  
    \caption{Charge density on the horizon after fine-tuning close to threshold via three different mechanisms: fine-tuning of $Q_0$ for the single-pulse, fine-tuning of $Q_0$ for the double-pulse, and fine-tuning of $\mathcal{A}_0$ for the single-pulse. For each of the three curves, the parameter in question has been fine-tuned to within a part in $10^{7}$, yielding horizon dynamics that have converged to the dynamical extremal regime for the range of advanced times plotted.} 
    \label{fig:horizonhair}
\end{figure}

The fact that the linearized Aretakis instability can imbue extremal black holes with some ``hair'' suggests that in the non-linear
case, there might be a related mild breaking of universality. In other words, not only are the proportionality constants in Eqs.~\ref{eq:Vtrapscaling}-\ref{eq:BHscaling2} family-dependent, but so are those in higher-order gradients of fields on the horizon. Indeed, the three sets of simulations discussed in this paper (the single pulse tuned via $Q_0$, the double-pulse tuned via $Q_0$, and the single-pulse tuned via $\mathcal{A}_0$) each yield different asymptotic values of the horizon charge density at extremality. This is shown in Figure~\ref{fig:horizonhair}.

Therefore, the horizon charge density (as well as the prefactor of the energy density growth) is a genuine ``hair" of the dynamical extremal black hole whose value is set independently of $Q$ and $M$. Such extremal hair has been discussed before on fixed metric backgrounds \cite{angelopoulos_horizon_2018,Bishoyi_hair,Aretakis_hair}, and it is interesting to see that it persists in the non-linear setting.

Turning back to Figure~\ref{fig:horizongrowth}, the growth of $T_{rr}$ along the extremal horizon is particularly surprising; whereas MRT \cite{murata_what_2013} saw \emph{constant} energy density on the horizon, growth should be much harder to achieve, as Einstein's equations should regulate any divergences in the stress tensor (modulo formation of a curvature singularity). However, in this case we find in our simulations that even though the components of the stress tensor blow up, the \emph{scalar} quantities that one can build out of the stress tensor (i.e. $T$ and $T_{\mu\nu}T^{\mu\nu}$) remain bounded along the dynamical extremal horizon. 
Indeed, the trace of the stress tensor for a spherically symmetric charged scalar field is\begin{align}
    T&=4f^{-1}\left[\left(\frac{P}{r}\right)_{,U}\left(\frac{P}{r}\right)_{,V}-\frac{Q_{,U}Q_{,V}}{64\pi^2\e^2P^2r^2}\right].
\end{align}
On the apparent horizon, $f^{-1}\partial_U\propto \partial_r$, so the divergent factor of $P_{,r}$ is balanced by the decaying factor of $P_{,V}$. In our simulations, we find that along the dynamical extremal event horizon,\begin{align}
    P_{,r}\sim V^{1/2}, \quad P_{,V}\sim V^{-3/2},
\end{align}
thus matching the scalings observed in the linear problem \cite{zimmerman_horizon_2017,Gelles_charge}. These power-law scalings ensure that $T\sim V^{-1}$. Physically, the decay of the stress tensor's trace indicates that even though the energy density diverges, it is balanced in the trace by an equal and opposite radial pressure and momentum density (as viewed in any timelike or null infalling frame).

This precise cancellation seems to ensure that all curvature scalars remain finite on the dynamical extremal horizon too. The trivial example is the Ricci scalar, being proportional to the trace of the stress tensor:\begin{align}
    R&=-8\pi T,
\end{align}
though we have confirmed the finiteness of the Kretschmann scalar, Weyl scalar, and Ricci tensor squared as well.

\begin{figure*}[t]
\centering
\begin{subfigure}
\centering
\includegraphics[width=0.99\textwidth]{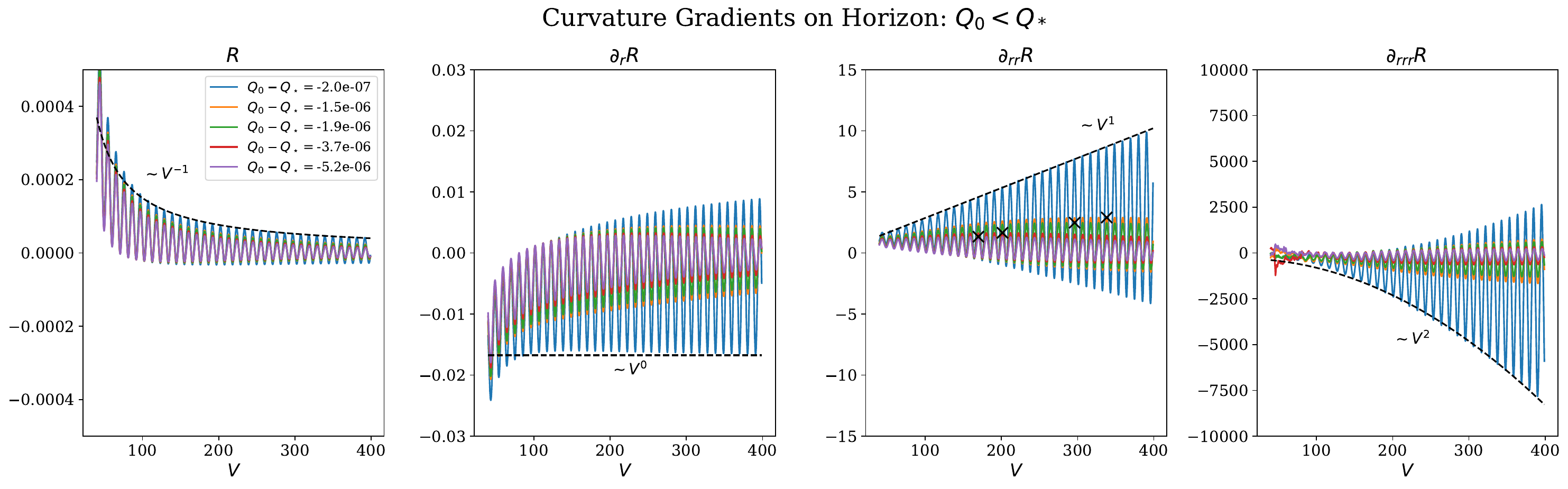}  
    \caption{Radial derivatives of the Ricci scalar along the outer horizon of a black hole as $Q_0\to Q_*$. The response of the Ricci scalar gradients at extremality is given by a power-law envelope bounding oscillations at the NZD frequency. Each radial derivative causes the exponent of the power-law envelope to increase by 1. The black crosses on the plot of $\p_{rr} R$ denote the location of $V_{\rm diss}$ --- the dissipation timescale of the instability. In Figure~\ref{fig:Vdissplot}, we show that $V_{\rm diss}$ scales with the critical exponent of $1/2$.} 
    \label{fig:curvaturegradient}
\end{subfigure}
\vspace{0.4cm}
\begin{subfigure}
\centering
\includegraphics[width=0.99\textwidth]{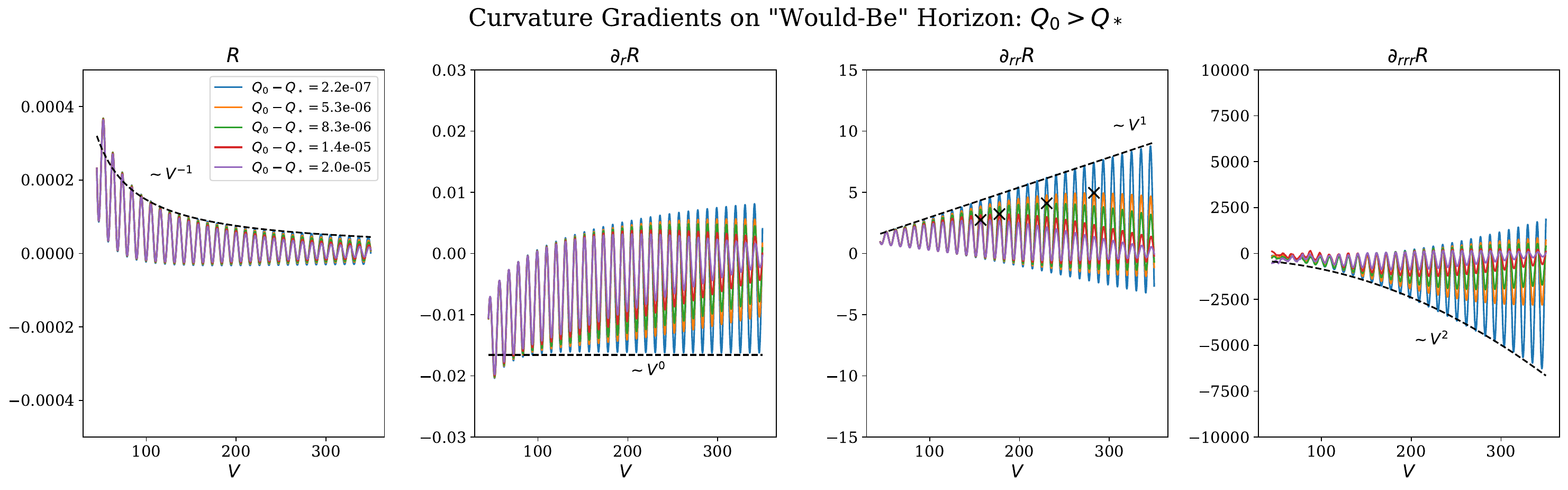}  
    \caption{Radial derivatives of the Ricci scalar along the ``would-be" horizon ($U=U_*$) of a super-extremal spacetime as $Q_0\to Q_*$ from above. The oscillatory and power-law response as we approach extremality from above matches the analogous approach from below in Figure~\ref{fig:curvaturegradient}. The black crosses on the plot of $\p_{rr} R$ denote the location of $V_{\rm diss}$ --- the dissipation timescale of the instability. In Figure~\ref{fig:Vdissplot}, we show that $V_{\rm diss}$ scales with the critical exponent of $1/2$.}
    \label{fig:curvaturegradient2}
\end{subfigure}
\label{fig:both2}
\end{figure*}

However, the radial derivatives of $T$ --- and hence the curvature \emph{gradients} (at sufficiently high order) --- diverge along the horizon. In Figure~\ref{fig:curvaturegradient}, we plot the Ricci scalar and its first three radial derivatives along the outer horizon of the black hole. We see that generically, the gradients grow up to a certain timescale $V_{\rm diss}$ before turning over and decaying again. This implies if we tune to exact extremality, $V_{\rm diss}\to\infty$ and higher-order gradients of curvature scalars grow without bound.

More specifically, the behavior of the curvature gradients is given by an oscillatory piece bounded by a power-law envelope. The exponents of the power-law envelope appear to satisfy\begin{align}
\label{eq:partialR}
   \p_r^n R|_{r_+}\sim V^{n-1},
\end{align}
which agrees with the scalings of $\p_r^n T$ that one can compute analytically from the expected blowup rates in the linear problem.

To measure the frequency of the oscillations in the curvature gradients, we repeat the procedure of our previous work \cite{Gelles_charge} and take a Fast Fourier Transform (FFT) of the data. We find that the FFT peaks at\begin{align}
    \frac{\omega_{\rm Ricci}}{\e }&=0.992 \pm 0.029,
\end{align}
where the error estimate comes from our Fourier-domain resolution: $\Delta\omega/\e=2\pi/\e V_{\rm max}=0.029$. That $\omega_{\rm Ricci}/\e\approx 1$ is no coincidence; as we showed in our linear paper \cite{Gelles_charge}, long-lived quasi-normal modes called Nearly-Zero-Damped (NZD) modes exist on a fixed near-extremal background, and the real part of this mode is precisely equal to $\e$ (see also \cite{hod_2010_v2,hod2012quasinormal,hod_2010_v2,yang2013branching,Zimmerman_2016_v2}). At exact extremality, the exponentially decaying NZD modes of sub-extremal black holes ``merge'' to give rise to power-law decay, and oscillations at the NZD frequency show up ubiquitously throughout the response of the matter fields on the fixed extremal horizon. Here, we see that these qualitative properties of the NZD mode persist in the non-linear problem, prompting the curvature gradients to oscillate while growing. In an upcoming paper, we will present a more detailed analysis of the NZD modes in the fully non-linear setting.

\begin{figure}[h]
    \centering
    \includegraphics[width=0.5\textwidth]{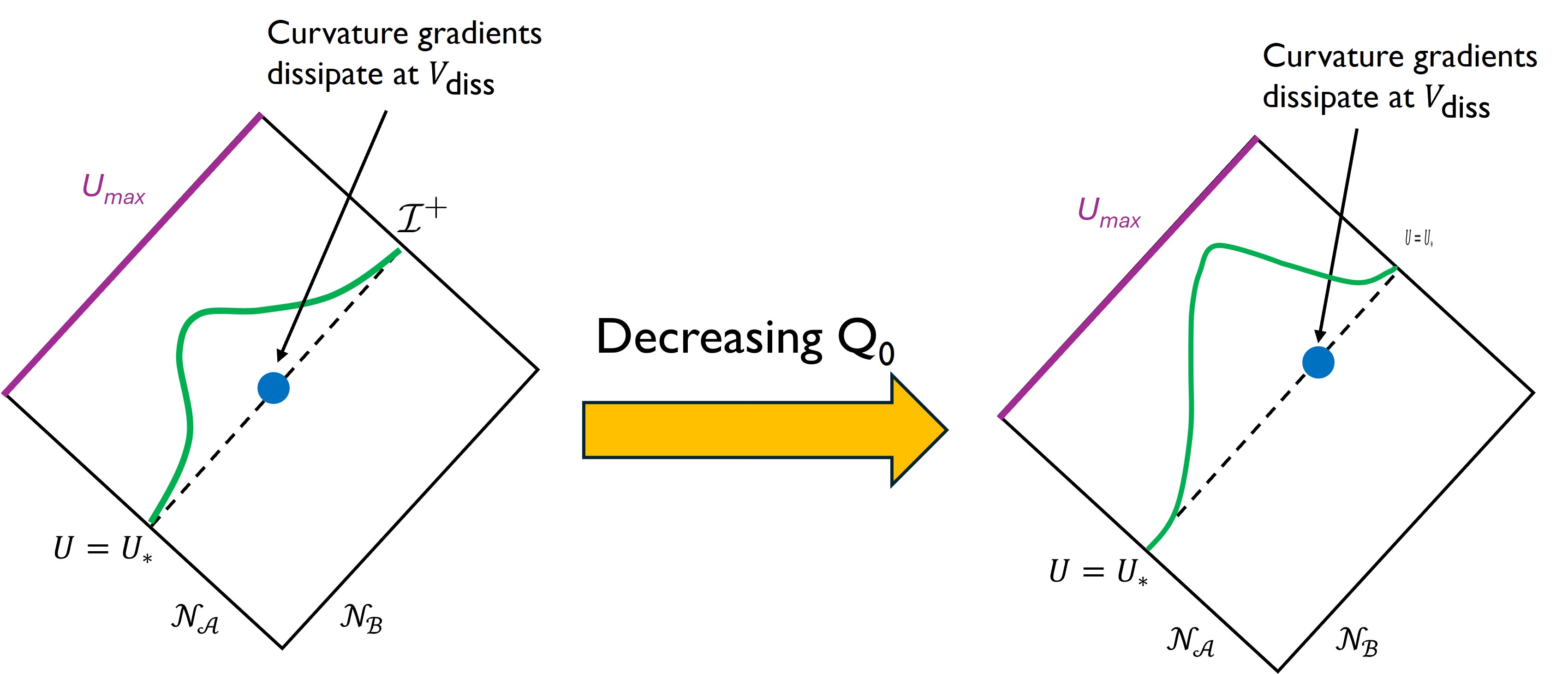}  
    \caption{Growth of large curvature gradients along a ``would-be horizon" at $U=U_*$. As $Q_0\to Q_*$ from above, the dissipation timescale $V_{\rm diss}$ is pushed to infinity. Prior to $V_{\rm diss}$, Aretakis-like growth occurs.} 
    \label{fig:subcritapproach}
\end{figure}

Now, it turns out this unstable behavior is seen on \emph{both} sides of the critical threshold: Even when there is no black hole, curvature gradients can still grow arbitrarily large. On this dispersive side of the threshold (i.e. $Q_0>Q_*$), we can quantify this behavior by tracking the curvature gradients along the hypersurface $U=U_*$, where $U_*$ is the time at which the extremal black hole would have formed. We can compute $U_*$ explicitly by extrapolating the $U$-coordinate of the sub-extremal horizons to exact extremality\footnote{Note that the results of this section not do change when we shift $U_*$ by small amounts, in particular within the uncertainties in estimating $U_*$.}. On the dispersive side of the threshold, the hypersurface $U=U_*$ (or any other $U={\rm const.}$ surface within our domain) is no longer an event horizon, so we refer to it as the ``would-be event horizon." The approach to $Q_0$ from the dispersive side of the threshold is shown as a cartoon in Figure~\ref{fig:subcritapproach}.

Employing this procedure, we compute curvature gradients along the would-be horizon and plot them in Figure~\ref{fig:curvaturegradient2}. There, we see the same growth rates and oscillation frequencies of the curvature gradients as we did in Figure~\ref{fig:curvaturegradient}. Thus, it does not matter whether you approach extremality from above or below: the Aretakis-like growth emerges with the same rates from either direction. Moreover, this growth occurs {\em before} a trapped region forms (i.e. $V<V_{\rm trap}$) on the $Q_0<Q_*$ side of the threshold. This indicates that the emergent Aretakis behavior is a property of a near dynamical extremal spacetime regardless of whether an event horizon is present within our domain of evolution.

Indeed, one can quantify deviation from extremality by analyzing the ``duration" of the instability on either side of the critical point. As previously mentioned, whenever $Q_0\neq Q_*$, the curvature gradients grow until an advanced time $V_{\rm diss}$ and then begin to decay. To be precise, let us define\footnote{We could have identified $V_{\rm diss}$ with any higher-order curvature gradient as well, but we chose to identify it with $\p_{rr}R$ since it is the lowest-order (and hence most ``well-behaved'' numerically) curvature gradient to exhibit genuine transient growth.}\begin{align}\label{eq:vdiss_def}
    V_{\rm diss}\equiv \text{$V$ at which $\p_{rr} R$ is maximized}.
\end{align}

\begin{figure}[h]
    \centering
    \includegraphics[width=0.5\textwidth]{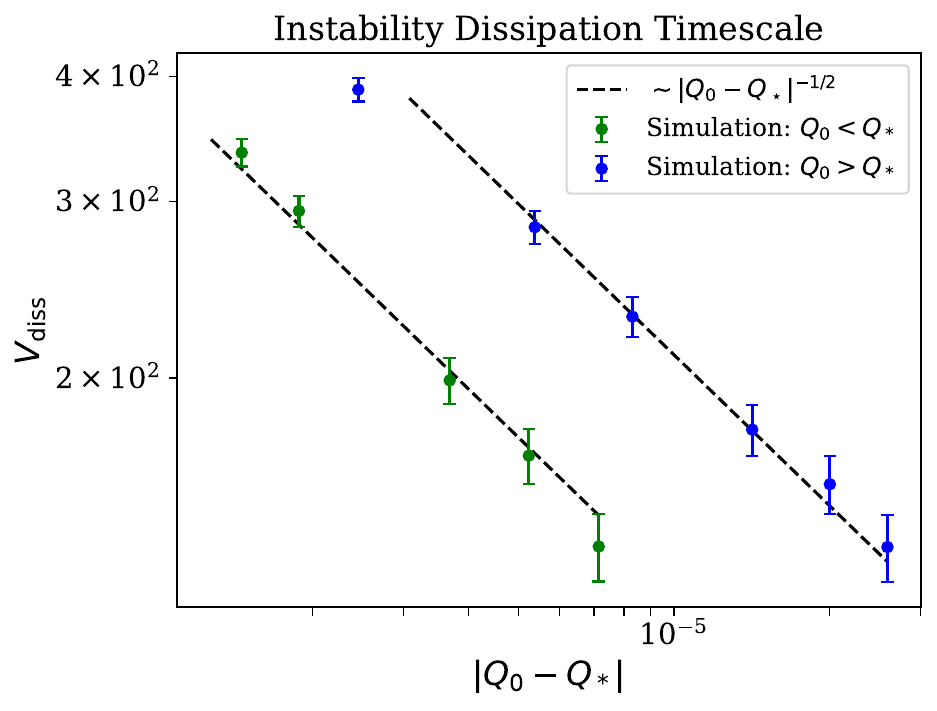}  
    \caption{The dissipation timescale $V_{\rm diss}$ (Eq.~\ref{eq:vdiss_def}) as the critical point $Q_*$ is approached for initial data consisting of the single-pulse. Error bars correspond to the period of the Nearly-Zero-Damped mode ($\approx 10$ for $\e Q_0=0.6$), which causes oscillations around the pure power-law. The critical scaling exponent of $1/2$ is the same on either side of the critical point.} 
    \label{fig:Vdissplot}
\end{figure}

Then plotting $V_{\rm diss}$ as a function of $Q_0-Q_*$ in Figure~\ref{fig:Vdissplot}, we see that the duration of the instability scales with the universal $1/2$ exponent \emph{on either side of the critical point}:\begin{align}
    V_{\rm diss}\sim |Q_0-Q_*|^{-1/2}.
\end{align}
Note that the pre-factor of the scaling law differs between the two sides of the critical point.

This matches the picture presented by Gralla, Zimmerman, \& Zimmerman \cite{gralla_transient_2016}, as well as Hadar \cite{hadar_near-extremal_2019}, both of whom derived analytically that transient instabilities of near-extremal horizons should last for a duration of $1/\kappa$, with $\kappa\equiv \frac{r_+-r_-}{2r_+^2}$ the outer surface gravity of the black hole. Near extremality,\begin{align}
    \kappa=\frac{r_+-r_-}{2r_+^2}\sim \sqrt{2(1-Q/M)},
\end{align}
which is consistent with the critical exponent of $1/2$. MRT \cite{murata_what_2013} found a similar relationship when analyzing the non-linear behavior of scalar fields incident on an extremal spacetime (for which backreaction causes the spacetime to become asymptotically sub-extremal). Our numerical results demonstrate that these predicted scalings carry over to the dispersive side of the critical point as well.

On the BH-forming side of the threshold, $V_{\rm diss}$ roughly coincides with $V_{\rm trap}$ --- the time at which a black hole forms and causes curvature gradients and energy density to decay along its apparent horizon. On the dispersive side of the threshold, $V_{\rm diss}$ coincides with the time at which $r(U_*,V)$ begins to turn up towards null infinity, similarly quenching the curvature gradients and energy density. The critical point $Q_0=Q_*$ corresponds to the in-between case: a black hole never forms along $U_*$ (in finite time), but outgoing null geodesics along $U_*$ never reach null infinity (in finite time).

In sum, the response of the matter fields on the dynamical extremal horizon mirrors that predicted by the linear theory. And furthermore, this picture extends cleanly to the dispersive side of the critical point, as curvature gradients (and hence the energy density) can grow without bound for sufficiently fine-tuned data along a would-be event horizon.

From here, we can examine what happens to this Aretakis-like growth in the fully non-linear setting as we move away from the (would-be) horizon. For the near-threshold solutions containing a black hole, this entails examining the matter fields in the black hole interior. For near-threshold solutions lacking a black hole, this entails examining the matter fields in the ``would-be interior" --- the region to the future of $U_*$ that would have been trapped had a black hole formed. On either side of the threshold, we find that the Aretakis-like divergence of gradients becomes more pronounced deeper in the interior, as we show below.

\subsection{(Would-Be) Interior}
In this subsection, we extend our analysis of dynamical extremal black holes to the interior of near-threshold solutions. 

To begin, let us follow MRT \cite{murata_what_2013} and consider a family of ingoing null geodesics $U(\lambda)$ parameterized by affine time $\lambda$. We can then examine quantities like energy density as a function of $\lambda$, which reveals how infalling null ``observers'' experience a near-threshold spacetime. 

\begin{figure}[h]
    \centering
    \includegraphics[width=0.5\textwidth]{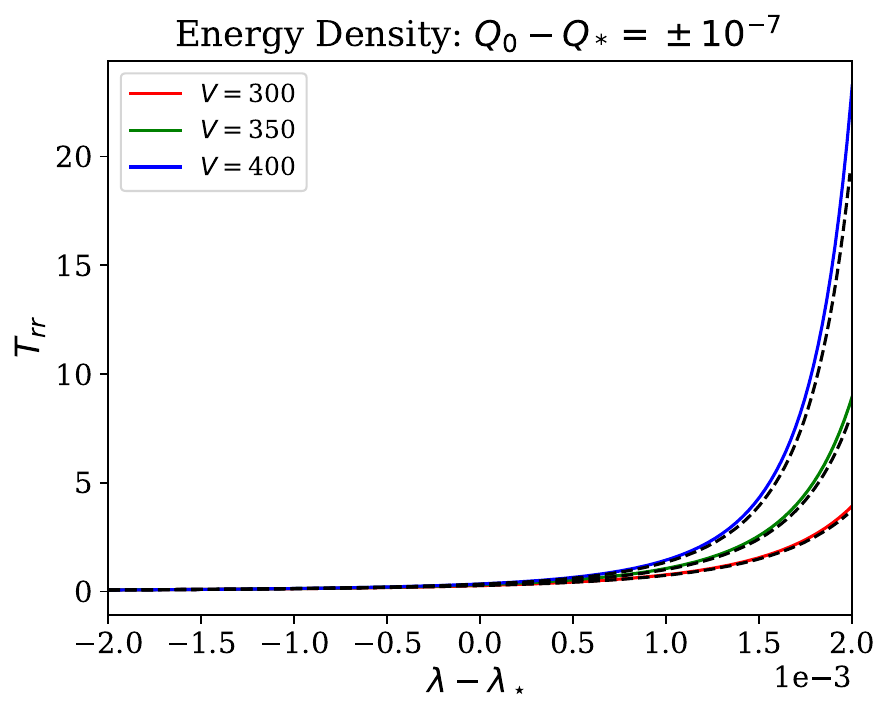}  
    \caption{Energy density plotted as a function of affine time along three ingoing null geodesics. The results are shown for a near-threshold solution on the BH-forming side of the critical point (solid lines) and dispersive side of the critical point (dashed lines).  While $T_{rr}$ begins to grow modestly on the (would-be) event horizon, the growth rate increases dramatically in the (would-be) interior. Generically, this trend persists for all ingoing null geodesics up to $V\sim V_{\rm diss}$, and this dissipation timescale can be made arbitrarily large by fine-tuning the initial data.} 
    \label{fig:Trrinterior}
\end{figure}

In Figure~\ref{fig:Trrinterior}, we plot the energy density (in the form of $T_{rr}$) as a function of $\Delta\lambda\equiv \lambda-\lambda_*$ for two near-threshold solutions: one black hole solution with $Q_*-Q_0=10^{-7}$, and one dispersive solution with $Q_*-Q_0=-10^{-7}$. For the BH-forming solution, $\lambda_*$ corresponds to the affine time at which the null geodesic crosses the event horizon into the interior. And for the dispersive solution, $\lambda_*$ corresponds to the affine time at which the null geodesic crosses the would-be event horizon into the would-be interior.

In this figure, as we cross the (would-be) event horizon from the exterior, we see dramatic growth as a function of affine time. Furthermore, this growth rate steepens as a function of $V$. The Aretakis-like behavior thus worsens as one enters the (would-be) interior. Importantly, the similarity of the solid and dashed lines in Figure~\ref{fig:Trrinterior} shows that this behavior is qualitatively the same on either side of the critical solution.

\begin{figure}[h]
    \centering
    \includegraphics[width=0.5\textwidth]{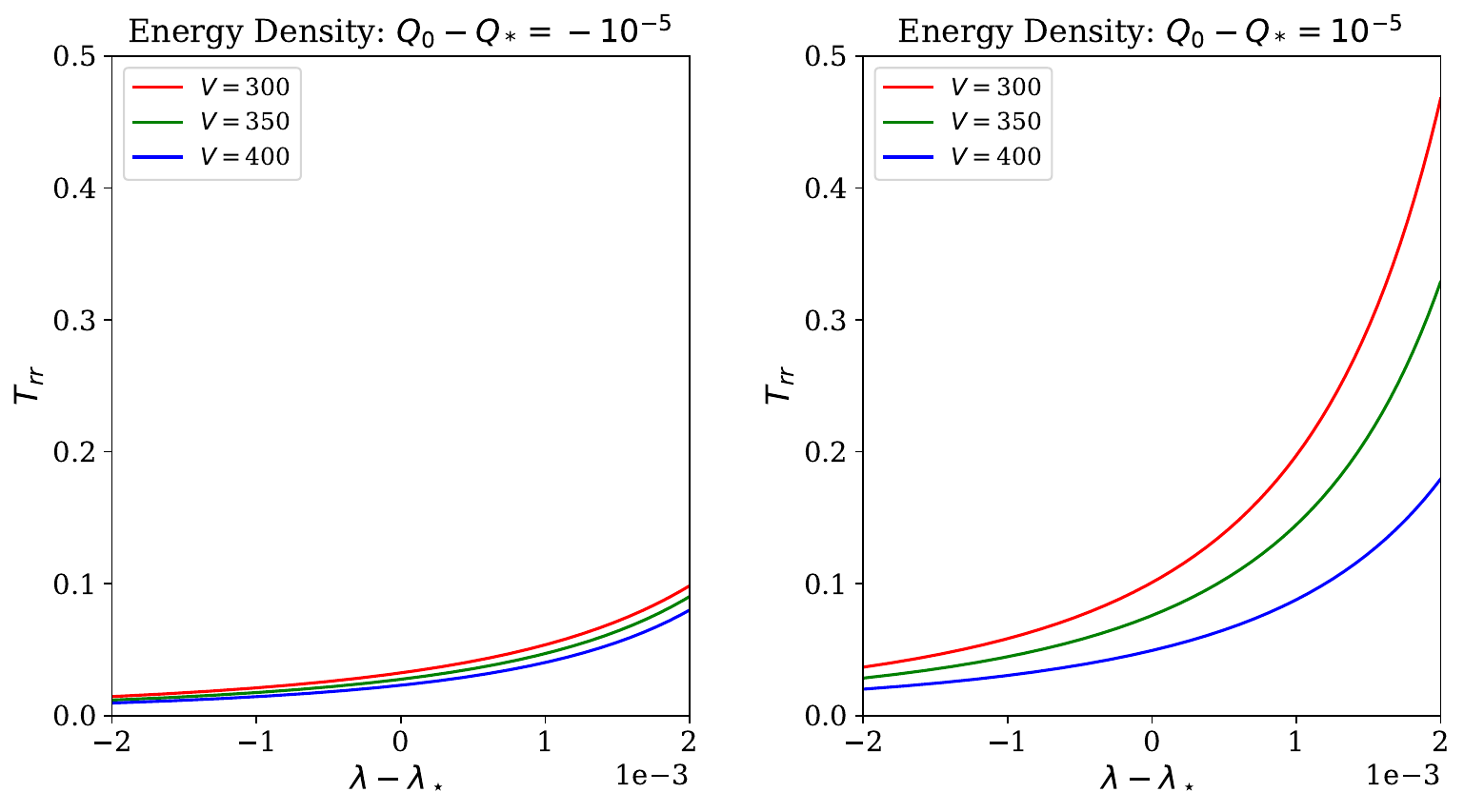}  
    \caption{Behavior of the energy density along three ingoing null geodesics, but {\em after} their corresponding dissipation timescales. Here, $V_{\rm diss}\approx 100$ for the simulation shown the left panel, and $V_{\rm diss}\approx 180$ for the simulation shown in the right panel. The growth with $V$ that is seen before $V_{\rm diss}$ (c.f. Figure~\ref{fig:Trrinterior}) has now become decay.} 
    \label{fig:dissinterior}
\end{figure}

Indeed, this divergence of the energy density appears to be a distinct property of the threshold solution. Departing from the threshold in either direction, the growth rate of the energy density in the (would-be) interior eventually begins to dissipate. This is demonstrated in Figure~\ref{fig:dissinterior}, for which we plot $T_{rr}$ as a function of affine time for two farther-from-threshold simulations than those shown in Figure~\ref{fig:Trrinterior}. In particular, Figure~\ref{fig:dissinterior} shows cases that we have evolved past their dissipation timescales $V_{\rm diss}$.

These numerical results are consistent with those of MRT $\cite{murata_what_2013}$, who also found that gradients diverge in the extremal interior. Our findings extend their work to the case of the charged scalar field and demonstrate that the instability of the extremal interior emerges on the dispersive side of the critical point too.

What is the physical origin of this instability? Drawing on previous work \cite{Marolf_extremal,Garfinkle_extreme,Marolf_Ori,murata_what_2013}, we find that both the Aretakis behavior on the dynamical extremal horizon and the subsequent divergence of energy density in the dynamical extremal interior can be explained in terms of geodesic focusing. We explain this phenomenon below, first summarizing the linear picture on a fixed background and then applying those results to our non-linear simulations.

\begin{figure*}[t]
    \centering
    \includegraphics[width=0.99\textwidth]{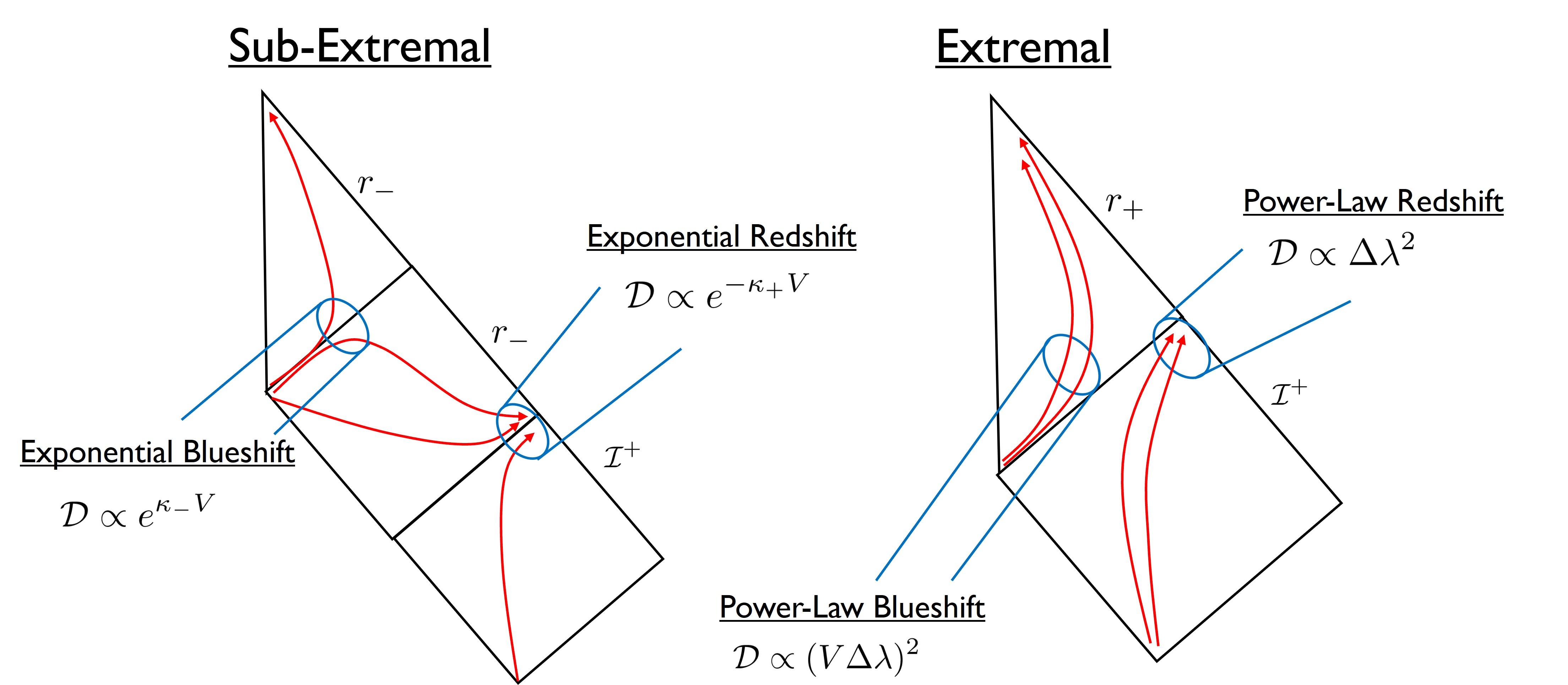}  
    \caption{Penrose diagram for static sub-extremal (left) and extremal Reissner-Nordstr\"om (right). Contours of constant $r$ are shown as red lines. In the sub-extremal case, the curves of constant $r$ exponentially cluster around the outer horizon but exponentially peel away from the inner horizon (Eq.~\ref{eq:exponentialstuff}). In the extremal limit, these two effects merge, creating a redshift in the exterior with a blueshift in the interior, and the exponentials break to power-laws (Eq.~\ref{eq:powerlawstuff}).}
    \label{fig:blueshiftcartoon}
\end{figure*}

\subsubsection{Focusing Instability: Linear Picture}
To understand the instability of the dynamical extremal interior, we will take a step back and analyze the behavior of matter perturbations on a fixed background. In particular, we will examine the behavior of outgoing radiation in the fixed RN interior. 

For static, sub-extremal Reissner-Nordstr\"om in Kruskal coordinates, curves of constant $r$ converge exponentially towards the outer horizon and diverge exponentially from the inner horizon:\begin{align}
\label{eq:exponentialstuff}
    \frac{\p U}{\p r}\bigg|_{r=r_+}\propto -e^{-\kappa_+ V},\qquad \frac{\p U}{\p r}\bigg|_{r=r_-}\propto -e^{\kappa_- V}. 
\end{align} 
Physically, one can think of the quantity $\p U/\p r$ tracking the frequency of outgoing radiation, as it measures the number of wavecrests ($dU$) packed into a radial displacement $(dr)$. This motivates us to define a ``Doppler factor" (in the heuristic sense):\begin{align}
\label{eq:Dopplerfac}
    \mathcal{D}\equiv \frac{\p U}{\p r}\left(\frac{\p U}{\p r}\bigg|_{V_{\rm Dop}}\right)^{-1},
\end{align} 
where the normalization term is evaluated on an arbitrary null slice $V=V_{\rm Dop}$ to ensure that $\mathcal{D}$ is invariant under re-parameterizations of $U$. Under this normalization, $\mathcal{D}$ directly measures the frequency shift of outgoing radiation emitted from $V_{\rm dop}$; when $\mathcal{D}<1$, the outgoing radiation has been redshifted, and when $\mathcal{D}>1$, the outgoing radiation has been blueshifted. So from the Doppler factors in Eq.~\ref{eq:exponentialstuff}, we see that outgoing radiation appears exponentially redshifted at the outer horizon and exponentially blueshifted at the inner horizon of a static, sub-extremal black hole. The exponential blueshift results in the Marolf-Ori shock \cite{Marolf_Ori}, which is a late-time instability of the sub-extremal inner horizon. 

On a fixed extremal background, however, the two horizons coalesce and the surface gravity vanishes, leading all Doppler effects to drop out completely as $\mathcal{D}\to 1$. This result is what ultimately allows Aretakis growth to persist at extremality: With no redshift, energy and charge can accumulate on the horizon without falling in.

But even though Doppler effects vanish \emph{precisely} along the fixed extremal horizon, they re-emerge everywhere else. Again taking $\Delta\lambda\equiv \lambda-\lambda_*$ to represent the affine time displacement from the moment of event horizon passage, we show in Appendix~\ref{app:extremalredshift} that on a fixed extremal background,\begin{align}
\label{eq:powerlawstuff}
     \mathcal{D}&\propto \begin{cases}
   \Delta \lambda^2,&\text{extremal exterior } 
    \\
        V^2\Delta \lambda^2,&\text{extremal interior }.
    \end{cases}
\end{align}
Both approximations are valid only sufficiently far from the extremal event horizon, in particular when $\Delta \lambda\gtrsim V^{-1}$. When $\Delta \lambda\lesssim V^{-1}$, we enter the near-horizon regime, for which the Doppler factor approaches 1. 

The quadratic dependence in Eq.~\ref{eq:powerlawstuff} indicates that an infalling observer sees outgoing radiation strongly blueshifted upon crossing into the interior of a static extremal black hole. This blueshift blows up at large $V$, resulting in what we call a focusing instability (to avoid confusion with the phrase ``blueshift instability,'' which is usually meant to refer to the Cauchy Horizon). This stands in contrast to the sub-extremal case, for which an observer must wait until crossing the inner horizon before seeing this blueshift. The rapid transition from redshift to blueshift upon crossing the fixed extremal horizon is depicted visually in Figure~\ref{fig:blueshiftcartoon}.


\subsubsection{Focusing Instability: Non-Linear Picture}
\label{sec:blueshiftnonlinear}
Here, we demonstrate that the focusing effects discussed for the fixed RN spacetimes are also present in the \emph{dynamical} extremal spacetimes, suggesting that one can think of these focusing effects as driving the growth of energy density (cf. Figures~\ref{fig:Trrinterior}-\ref{fig:dissinterior}) and curvature gradients. 

\begin{figure}[h]
    \centering
    \includegraphics[width=0.4\textwidth]{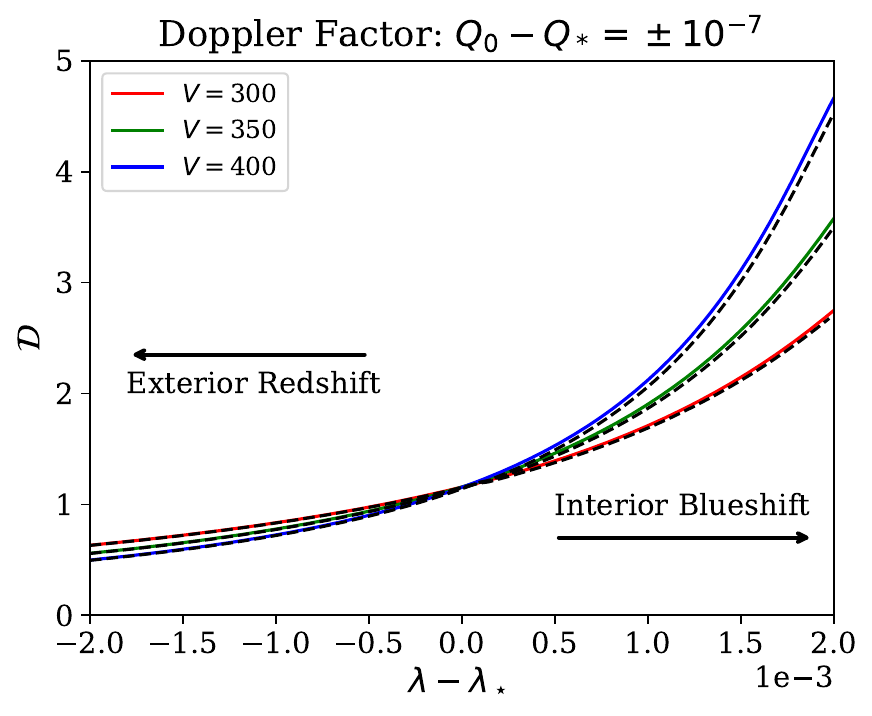}  
    \caption{Doppler factor (defined in Eq.~\ref{eq:Dopplerfac}) plotted as a function of affine time along several ingoing null geodesics. The results are shown for a near-threshold solution on the BH-forming side of the critical point (solid lines) and dispersive side of the critical point (dashed lines).  We see that the outgoing radiation is redshifted going toward the outside of the (would-be) horizon but becomes rapidly blueshifted moving into the (would-be) interior. } 
    \label{fig:Doppler}
\end{figure}

In Figure~\ref{fig:Doppler}, we plot the Doppler factors from the fully non-linear, near-threshold simulations. We take $V_{\rm dop}=100$ to minimize the effects of early-time transients. In this figure, we see that around the vicinity of the (would-be) horizon, there is no appreciable Doppler shift, whereas the (would-be) interior displays a growing blueshift. This blueshift grows both with time $V$ and distance into the interior. This transition from redshift to blueshift is also quite sharp (note the horizontal axis scale), and the trends suggest that $\mathcal{D}\to 0$ for $\lambda-\lambda_*\lesssim 0$ and $\mathcal{D}\to \infty$ for $\lambda-\lambda_*\gtrsim0$ as $V\rightarrow\infty$. 

Moreover, the solid and dashed curves qualitatively agree in Figure~\ref{fig:Doppler}, indicating that the Doppler factors of the near-threshold solutions do not depend on whether a black hole is actually present, and the focusing of interior, outgoing radiation begins even in the absence of trapped surfaces. This is consistent with the results in \S\ref{sec:aretakis}, where we showed that the dynamics along the hypersurface $U=U_*$ behave like an extremal horizon up until a dissipation timescale $V_{\rm diss}\sim (Q_0-Q_*)^{-1/2}$. So for all $V\lesssim V_{\rm diss}$, we expect contours of constant $r$ to cluster about the would-be horizon from the outside, creating a redshift. Conversely, contours of constant $r$ should peel away from the would-be horizon towards the inside, creating a blueshift. 

\begin{figure}[h]
    \centering
    \includegraphics[width=0.45\textwidth]{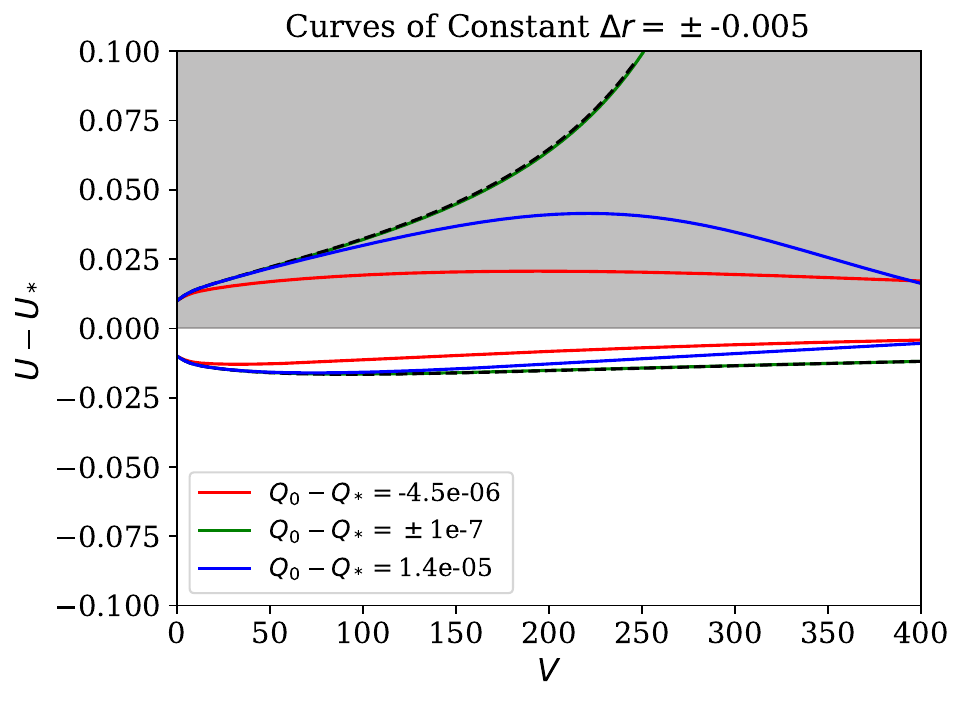}  
    \caption{Segment of the (rotated) Penrose diagrams of several dynamical spacetimes, showing two contours of constant $\Delta r$ for each, where $\Delta r$ is defined as the radial displacement from the (would-be) horizon at a fixed $V$. The (would-be) interior is shown in the top half of the plot and is shaded in light gray. Curves that move away from (toward) $U=U_*$ indicate blueshifting (redshifting). The green curve is from the nearest-to-threshold solution on the BH-forming side of the critical point, while the black dashed line is from the nearest-to-threshold solution on the dispersive side of the critical point.} 
    \label{fig:wouldbeinterior}
\end{figure}

This is qualitatively consistent with our numerical evolutions. In Figure~\ref{fig:wouldbeinterior}, we plot contours of constant $\Delta r=\pm 0.005$ in a segment of the (rotated) Penrose diagram for four different simulations (two near-threshold, two farther-from-threshold). This illustrates that in the interior, at early times (relative to $V_{\rm diss}$), the behavior of the curves is consistent with a growing blueshift. Conversely, in the exterior, we see contours consistent with steady redshifting. In this figure, $V_{\rm diss}$ lies between 100 and 200 for the farther-from-threshold solutions, while $V_{\rm diss}$ exceeds 400 for the near-threshold solutions. 

We note that in this figure, the contour of constant $\Delta r=-0.005$ (interior) extends to larger $U$ on the dispersive side of the threshold (blue curve) than it does on the BH-forming side of the threshold (red curve). This is because the BH-forming solution contains an \emph{inner} horizon --- in this case forming at $U-U_*=0.035$ and $V_{\rm trap}\approx 250$ --- that alters the range of $U$ accessible to contours of constant $\Delta r$.


This focusing picture is largely the same as the one presented by Marolf \& Ori \cite{Marolf_Ori} for sub-extremal black holes, except here we have repeated their calculations precisely at extremality. 
Also, an analogous focusing instability operates at the sub-extremal Cauchy Horizon (as originally established by Penrose \cite{Penrose_blueshift} and proven rigorously by Dafermos \cite{Dafermos:2003vim}). This Cauchy Horizon instability --- which arises from the exponential blueshifting of \emph{ingoing} radiation --- causes so-called mass inflation, but only in sub-extremal black holes \cite{Poisson_massinflation,murata_what_2013,gajic2019interior}. In contrast, the focusing instability we have discussed here arises from a power law blueshifting of \emph{outgoing} radiation near the inside of the (would-be) event horizon, and it persists at (fine-tuned dynamical) extremality in the non-linear regime. 

In the next section we speculate about the endpoint of this focusing instability --- and whether or not it might form a curvature singularity.

\subsection{Singularity}

Our results indicate that the focusing of outgoing radiation acts as the primary physical mechanism causing the energy density to diverge at threshold. This growth of the energy density does not seem to be restricted to an effective thin shell on the horizon, but rather extends (at least some distance) into the interior. Given this picture, it is difficult to see how an interior curvature singularity would {\em not} form as $V\rightarrow\infty$. For near-threshold solutions that form black holes, this putative unbounded curvature would be censored from external view by an event horizon. Presumably, however, this unbounded curvature would be \emph{uncensored} on the dispersive side of the threshold; since 
outgoing geodesics in some portion of the would-be interior reach future null infinity, observers there would be able to ``see" this region of arbitrarily large curvature. 

A significant caveat with this conjecture --- which we cannot address here --- relates to the question raised earlier of what volume of the interior is universal. The growth that we see in the interior might be an artifact of our background spacetime being super-extremal, singular RN, rather than a feature of the putative universal threshold solution. Indeed, on our ingoing initial data slice, $(U,V)=(U_{\rm max},0)$ corresponds to the RN singularity at $r=0$, and hence $U=U_{\rm max}$ is effectively a Cauchy horizon for this class of spacetimes. We are not able to numerically evolve all the way to $U_{\rm max}$, but outgoing null rays ($U={\rm const.}$) that are close to $U_{\rm max}$ are focused into an ever-narrowing sliver of proper radius close to the would-be horizon\footnote{This logic could be used to argue that we are only seeing an effective thin shell of divergent energy density in the limit, which one can imagine is less likely to result in a curvature singularity. However, the fact that we do not see any turn-over in energy density growth as one evolves inward --- as one would expect for a shell-like distribution of energy --- suggests otherwise.}. 

\begin{figure}[h]
    \centering
    \includegraphics[width=0.5\textwidth]{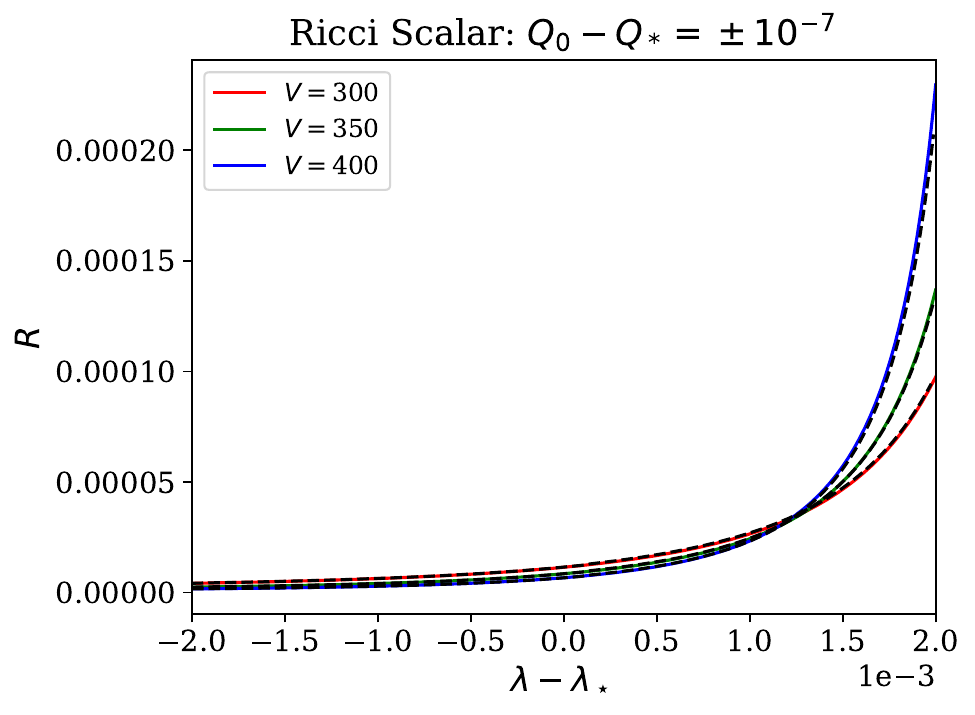}  
    \caption{Behavior of the Ricci scalar in the near-threshold solutions. Since the Ricci scalar oscillates strongly at the NZD mode (see Figure~\ref{fig:curvaturegradient}), the curves here have been averaged over one NZD period in the $V$ direction.} 
    \label{fig:Riccifig}
\end{figure}

With these caveats in mind, in Figure~\ref{fig:Riccifig} we plot the Ricci scalar $R$ as a function of affine time about the (would-be) horizon for near-threshold solutions. Note that at the horizon ($\lambda=\lambda_*$), we see slow decay of $R$, as expected from the Aretakis-like scaling discussed in \S\ref{sec:aretakis}. But moving into the interior, this trend rapidly changes (again notice the horizontal axis scale), and we start seeing significant growth with $V$. This suggests a curvature singularity may indeed form asymptotically at threshold.



\section{Conclusion}
\label{sec:conclusion}
In this work, we have numerically explored dynamical extremal spacetimes formed using charged scalar fields. By scattering wavepackets off of a super-extremal spacetime and fine tuning one-parameter families of initial data, we can construct extremal threshold solutions.

Extending the analysis of MRT \cite{murata_what_2013}, our work incorporates the effects of electromagnetic coupling to the matter and focuses on the \emph{approach} to the threshold solution. We have analyzed both sides of the critical point: \begin{itemize}
    \item The BH-forming side (e.g. $Q_0<Q_*$), for which a trapped surface forms at a large advanced time
    \item The dispersive side (e.g. $Q_0>Q_*$), for which the outgoing null expansion becomes small on a ``would-be horizon" but never actually reaches zero.
\end{itemize}

The threshold solutions appear to be universal, modulo a family-dependent horizon charge density. Near-critical solutions exhibit properties governed by universal scaling laws relative to the distance from threshold in parameter space. For example, the time $V_c$ during which the solution remains close to the universal solution scales as $V_c \propto |P-P_*|^{-1/2}$. Here, $P$ refers to any one-parameter family of initial data that smoothly crosses the threshold at $P=P_*$. On the BH-forming side of the threshold, $V_c\approx V_{\rm trap}$ --- the time at which a trapped surface forms. On both sides of the threshold, $V_c\approx V_{\rm diss}$ --- the time at which scalar field and curvature gradients stop growing along the (would-be) horizon. This is the kind of scaling seen in so-called Type I critical collapse~\cite{gundlach2007critical,Gundlach:2025yje}. Note that in Appendix~\ref{app:critfit}, we explore potential deviations from the $1/2$ critical scaling of $V_{\rm trap}$, and future work evolving data to larger $V$ will enable more precise constraints.

Aretakis-like growth of energy density emerges near threshold, and this growth continues into the interior --- whether that be the actual black hole interior on the BH-forming side of threshold, or the would-be interior on the other side. We re-emphasize that in the former case, this growth can occur {\em before} any trapped surfaces form. We demonstrate that these large gradients can be explained physically in terms of the focusing of outgoing radiation, as the (near-)extremal horizon separates an exterior redshift from a strong interior blueshift. This qualitative picture unifies insights from the past works of Marolf \cite{Marolf_extremal}, Garfinkle \cite{Garfinkle_extreme}, Marolf \& Ori \cite{Marolf_Ori}, and MRT \cite{murata_what_2013}.

The Aretakis-like growth and eventual divergence of energy density does {\em not} seem to result in a curvature singularity on the horizon, as a careful balance between energy density, radial pressure, and momentum density leaves all curvature scalars finite there. But this cancellation does not seem to happen inside the event horizon, and we have given evidence that curvature asymptotically diverges in the interior. Such growth of curvature will be censored from external view on the BH-forming side of threshold, but apparently not so on the dispersive side. By finely tuning the initial data, one can make $V_{\rm diss}$ arbitrarily large and hence initialize this divergence mechanism an arbitrarily short proper distance inside the (would-be) event horizon. However, for a physical scenario with finite $V_{\rm diss}$, divergences will all eventually turn over in $V$.

In order to determine the genericity of the focusing singularity within the ``sub-manifold'' \cite{angelopoulos2024nonlinear} of dynamical extremal spacetimes, we will need to determine what region of the interior is in fact universal. If the universal portion of the interior does {\em not} extend into the region with significant blueshifting (i.e $\Delta \lambda\gtrsim V^{-1}$ from Eq.~\ref{eq:powerlawstuff}), then the apparent singularity formation we see could merely be an artifact of the super-extremal RN spacetime on our ingoing hypersurface $\mathcal{N}_A$. Thus, important follow-up work to address this question will involve specifying appropriate outgoing radiation on this surface. The ideal scenario would be to ``regularize'' the interior by pasting an outgoing pulse of matter on $\mathcal{N}_A$; in principle, such a pulse could cause the metric on $\mathcal{N}_A$ to become flat at $r=0$, or it could cause a small black hole to form on $\mathcal{N}_A$ for some $r\ll 1$. 

In this procedure, it will be important to analyze the role of the coupling constant $\e Q_0$. While we have so far restricted our analysis to the fixed dimensionless coupling $\e Q_0=0.6$, an outgoing perturbation's ability to regularize the RN singularity will likely depend on the coupling strength of the field. Even the dynamics of the ingoing perturbations alone may depend on $\e Q_0$ in a way that goes beyond the linear picture, and we hope to explore this in future work.

It will also be interesting to investigate universality within the broader context of different matter models, eventually including gravitational radiation by going beyond spherical symmetry. In critical gravitational collapse, it is well known that the type of threshold solution (stationary Type I or self-similar Type II), as well as the particular values of the scaling exponents both depend on the kind of matter/energy undergoing collapse~\cite{gundlach2007critical,Gundlach:2025yje}. It is therefore quite interesting that for the known extremal threshold cases --- uncharged scalar fields \cite{murata_what_2013}, charged scalar fields as described here, and collapsing Vlasov matter described in ~\cite{kehle2024extremal,east2025gravitational} --- the same scaling exponent of $1/2$ governs near-threshold dynamics (at least with regard to certain parameters). Moreover, with charged scalar field collapse in spherical symmetry beginning from regular initial data, threshold solutions have been found that exhibit Type II self-similar collapse, with near-threshold BH-forming solutions approaching Schwarzschild and not RN \cite{Gundlach:1996vv,Hod_1997_crit,Petryk2005}. 
Thus, ``universality'' is certainly more complex and nuanced than what some anticipated following Choptuik's discovery.

A related question concerns the ``relative stability'' of threshold solutions when multiple forms of energy are present~\cite{choptuik_rs,Gundlach:2019wnk}. For example, given that the class of extremal threshold solutions of Vlasov collapse found in~\cite{kehle2024extremal,east2025gravitational} are exactly RN outside of the Vlasov distribution, then these critical solutions should be Aretakis-unstable to scalar field perturbations (at least at the linear level). It would be interesting to see what the generic non-linear end state of threshold solutions are in this Vlasov-plus-(charged)-scalar field scenario. 


\section*{Acknowledgements}
We thank Daniela Cors, Mihalis Dafermos, Dejan Gajic, David Garfinkle, Delilah Gates, Elena Giorgi, Carsten Gundlach, Lam Hui, Thomas M\"adler, Laetitia Martel, Christoph Kehle, Amos Ori, Harvey Reall, Ryan Unger, Maxime van de Moortel, Achilleas Porfyriadis, and Noa Zilberman for many helpful discussions. ZG was supported by a NSF Graduate Research Fellowship and benefited from computing resources at Princeton University. FP acknowledges support from the NSF through the grants PHY-220728 and
PHY-2512075.

\appendix
\begin{widetext}
\section{Super-Extremal Initial Conditions}
\label{app:initdata}
The Einstein constraint equations in double null coordinates (Eqs.~\ref{eq:cons1}-\ref{eq:cons2}) admit analytic solutions:\begin{align}
\label{eq:conssol}
    f(U,V_0)|_{\mathcal{N}_A}&=f_A(U,V_0)e^{\int_{U_0}^U dU'\frac{8\pi r(U',V_0)|D_U\phi(U',V_0)|^2}{r_{,U}(U',V_0)}}, \qquad
    f(U_0,V)|_{\mathcal{N}_B}=f_B(U_0,V)e^{\int_{V_0}^V dV'\frac{8\pi r(U_0,V')|D_V\phi(U_0,V')|^2}{r_{,V}(U_0,V')}},
\end{align}   
where\begin{align}
   f_A(U,V_0)&=\frac{f(U_0,V_0) r_{,U}(U,V_0)}{r_{,U}(U_0,V_0)}, \qquad f_{B}(U_0,V)=\frac{f(U_0,V_0) r_{,V}(U_0,V)}{r_{,V}(U_0,V_0)}.
\end{align}

Now, if the initial data is extremal or sub-extremal, then one can simply take $r$ to match its Reissner-Nordstr\"om value in MRT gauge\footnote{Note that in our linear paper \cite{Gelles_charge}, we rescaled $f$ by a factor of $2$ from MRT so that $f$ agrees with the $UV$ metric component. We analogously rescaled the argument of $r_*$ in Eq.~\ref{eq:MRTgauge}, thus altering $r_{,U}$ by a factor of 2 compared to MRT as well.}, and then plug this quantity into the above expressions for $f$. Recall that MRT gauge sets $r(U,V)$ via the correspondence\begin{align}
\label{eq:MRTgauge}
    r_\star(r_++V/2)+r_\star(r_+-U/2)=r_\star(r).
\end{align}
In our previous work, we computed both $r$ and its derivatives in this gauge, from which $f$ can be subsequently computed via Eq.~\ref{eq:conssol}.

However, MRT gauge is not adapted to super-extremal spacetimes since Eq.~\ref{eq:MRTgauge} contains factors of $r_+$. So in the case of a super-extremal background, we follow \cite{murata_what_2013} and take $r(U,V)$ to just match its extremal value along the initial data hypersurfaces (i.e. solve Eq.~\ref{eq:MRTgauge} with $r_+=1$). We call this choice ``extremal gauge."

To compute $f_{\rm ex}$ --- the $UV$ metric component in extremal gauge --- we start by requiring that $M(U_0,V_0)=1$, giving\begin{align}
     1-\frac{2M}{r}+\frac{Q^2}{r^2}=-\frac{2r_{,U}r_{,V}}{f} \Longrightarrow f_{\rm ex}(U_0,V_0)=-\frac{2r^2r_{,U}r_{,V}}{Q^2+r(r-2)}\bigg|_{(U_0,V_0)}.
\end{align}
Evaluating $r_{,U}$ and $r_{,V}$ then gives us the expressions for $f_{A}$ and $f_B$ in extremal gauge:\begin{align}
    f_{A,\rm ex}=\frac{F(r)}{2F(1+V_0/2)F(1-U/2)}\frac{1+r_0(r_0-2)}{Q_0^2+r_0(r_0-2)},\quad f_{B,\rm ex}=\frac{F(r)}{2F(1+V/2)F(1-U_0/2)}\frac{1+r_0(r_0-2)}{Q_0^2+r_0(r_0-2)},
\end{align}
where $F(r)\equiv (r-1)^2/r^2$. These can then be plugged into Eq.~\ref{eq:conssol} to give the fully dynamical $f_{\rm ex}$.

\section{Fitting the Critical Exponents}
\label{app:critfit}
In this Appendix, we provide a more detailed analysis of the scaling of $V_{\rm trap}$. As explained in \S\ref{sec:results}, our simulation results indicate that $V_{\rm trap}\to\infty$ near some critical point $Q_*$, suggesting that $V_{\rm trap}$ can be fit to a power-law of the form\begin{align}
    V_{\rm trap}\sim |Q-Q_*|^{-p}
\end{align}
for $p>0$. However, both $p$ and $Q_*$ are unknowns, and the two are highly degenerate. Since we have been able to evolve only to $V_{\rm max}=400$, we are limited in our ability to properly constrain each of these parameters.

Nonetheless, we can still perform an explicit fit to our data. In particular, for each family of initial data, we perform a joint least-squares fit (in log-log space) of $V_{\rm trap}$, $r/M$, and $Q/M$ to the functional forms:\begin{align}
\label{eq:fitpars}
    V_{\rm trap}&=C_V(Q_*-Q_0)^{-p_V},\qquad r/M-1=C_r(Q_*-Q_0)^{p_r},\qquad 1-Q/M=C_Q(Q_*-Q_0)^{p_Q}.
\end{align}
This gives seven unknowns in total: $\{C_V,C_r,C_Q,p_V,p_r,p_Q,Q_*\}$. Performing the fit jointly ensures that $V_{\rm trap},\,r/M,$ and $Q/M$ all share the same critical point $Q_*$ (or $\mathcal{A}_*$ for the ``vary-amplitude" family of initial data). 

Now in performing this fit, the predominant source of uncertainty comes from the range of data points chosen; when $V_{\rm trap}$ lies within the pulse, transient effects dominate, but when $V_{\rm trap}$ lies near the outer edge of the simulation domain, the reliableness of the power law fit breaks down. This leaves an intermediate range of points from which we should ideally sample, but the endpoints of this range are not exactly known.

Therefore, we can estimate the uncertainty on the seven inferred parameters (and in particular on $p$) by repeating the fit procedure with several different ``windows" of data points. In particular, we fit Eq.~\ref{eq:fitpars} for all possible windows of points that satisfy $20\leq V_{\rm trap,min}\leq 150$  and $150\leq V_{\rm trap,max}\leq 390$, subject to the constraint that each window contain at least five data points. For the three different initial data families, this procedure gives the following spreads in $p$ under the different fitting windows:\begin{align}
    \underline{\text{Single-Pulse}}:& \quad p_V\in[0.440, 0.799],\quad p_r\in[0.465,0.531],\quad p_Q\in[0.932,1.062]
    \\
    \underline{\text{Double-Pulse}}:& \quad p_V\in[0.438, 0.613],\quad p_r\in[0.464,0.568],\quad p_Q\in[0.931,1.133]
    \\
    \underline{\text{Vary-Amplitude}}:& \quad p_V\in[0.481, 0.705],\quad p_r\in[0.492,0.535],\quad p_Q\in[0.984,1.07].
\end{align}

All ranges are consistent with the theoretically motivated $\{p_V=1/2,p_r=1/2,p_Q=1\}$ critical exponents. However, the spread in $p_r$ and $p_Q$ is significantly tighter than it is in $p_V$, reflecting the fact that we have not yet explored a sufficiently large range in $\ln(Q_*-Q_0)$ to constrain $p_V$ to a high degree of precision. Future work will focus on improving the computational tools needed to efficiently evolve to larger $V$ and improve those constraints.

\section{Extremal Redshift Factors}
\label{app:extremalredshift}
In fixed Reissner-Nordstr\"om in MRT gauge (and with $f$ normalized as in our first paper \cite{Gelles_charge}), the radial coordinate satisfies \begin{align}
\label{eq:rUeq}
   r_{,U}&=\frac{F(r)}{2F(r_+-U/2)} ,\qquad r_{,V}=\frac{F(r)}{2F(r_++V/2)}.
\end{align}
We would like to evaluate the first expression at extremality in the limit of large $V$. This will require us to expand $r(U,V)$ around $V=\infty$.

Now, this expansion is qualitatively different outside the black hole $(r>r_+)$ versus inside the black hole $(r<r_+)$. Outside the black hole, $r(U,V)\to \infty$ for large $V$, whereas inside the black hole $r(U,V)\to r_+$ for large $V$. Therefore, we parameterize the series expansions as \begin{align}
    r(U,V)|_{r>r_+}&=A_{0,\rm ext}V+B_{0,\rm ext}+\mathcal{O}(V^{-1})
    \\
    r(U,V)|_{r<r_+}&=r_+-\frac{A_{0,\rm int}}{V}+\frac{B_{0,\rm int}\log V+C_{0,\rm int}}{V^2}+\mathcal{O}\left(\frac{(\log V)^2}{V^3}\right).
\end{align}
As we will see, the factors of $\log V$ in the above equation will be necessary for a consistent expansion in the interior.

From here, we take these expressions and plug them into Eq.~\ref{eq:MRTgauge}, expanding in large $V$ and equating like terms to find\begin{align}
\label{eq:coeffs}
    A_{0,\rm ext}&=\frac{1}{2},\quad B_{0,\rm ext}=r_++\frac{2r_+^2}{U}-\frac{U}{2}+r_+\log\left(\frac{U^2}{4r_+^2}\right),
    \\
    A_{0,\rm int}&=2r_+^2,\quad B_{0,\rm int}=16r_+^3,\quad C_{0,\rm int}=\frac{8r_+^4}{U}-2r_+^2U+8r_+^3\log\left(\frac{U}{8r_+}\right),
\end{align}
where $U=0$ corresponds to the extremal horizon. Now, these approximations hold only when the leading-order term dominates over the sub-leading terms, which happens when\begin{align}
    |A_{0,\rm ext}V|>|B_{0,\rm ext}|\Leftrightarrow |U|\gtrsim V^{-1},\qquad |A_{0,\rm int}|>\left|\frac{B_{0,\rm int}\log V+C_{0,\rm int}}{V^2}\right|\Leftrightarrow |U|\gtrsim V^{-1},
\end{align}
Here, we have assumed large $V$ and ignored order unity pre-factors. Thus, the ``large-$V$" regime specifically means that $V\gtrsim \Delta U^{-1}$, with $\Delta U$ the null displacement from the horizon.

Now, plugging the coefficients in Eq.~\ref{eq:coeffs} into the expression for $r_{,U}$ then gives\begin{align}
\label{eq:Uexpressions}
    \frac{\p U}{\p r}&= \begin{cases}
    \frac{2\Delta U^2}{(\Delta U-2r_+)^2}+\mathcal{O}(V^{-1}),&{\rm exterior}
    \\
        \frac{(V\Delta U)^2}{2r_+^2(2r_+-\Delta U)^2}+\mathcal{O}(V\log V),&{\rm interior}
    \end{cases}
\end{align}
from which we obtain the leading-order scalings in Eq.~\ref{eq:powerlawstuff}.

Finally, since the above expressions hold only in MRT gauge, it is worth converting $U$ to a more physically invariant variable. To that end, we follow MRT \cite{murata_what_2013} and express $U$ in terms of the affine time along an ingoing null geodesic $U(\lambda)$. To find $\lambda$, we solve the geodesic equation in double-null coordinates:\begin{align}
    \frac{d^2U}{d\lambda^2}+\left(\frac{dU}{d\lambda}\right)^2\frac{f_{,U}}{f}=0\Longrightarrow \lambda\propto \int^U dU'f(U',V).
\end{align}
One can fix the lower-bound of the integral to live at $U_{\rm EH}$ (so that $\lambda=0$ at the event horizon), and one can set the proportionality constant by requiring that each geodesic be released with unit velocity from $\mathcal{N}_B$:\begin{align}
    \lambda\equiv \frac{\int_{U_{\rm EH}}^UdU'f(U',V)}{f(U_0,V)}.
\end{align}
Using this definition and plugging in $f=\frac{1}{2}$ in MRT gauge on the fixed extremal RN background, we obtain the scaling $U\sim \lambda$ up to an order-unity prefactor. Substituting $U\to\lambda$ into Eq.~\ref{eq:Uexpressions} then gives Eq.~\ref{eq:powerlawstuff}.

\end{widetext}

\bibliographystyle{apsrev4-2} 
\bibliography{bib,references}
\end{document}